\documentclass[
  journal=pasa,
  manuscript=research-article,
  year=2025,
  volume=37,
]{cup-journal}

\DeclareRobustCommand{\VAN}[3]{#2}
\let\VANthebibliography\thebibliography
\def\thebibliography{\DeclareRobustCommand{\VAN}[3]{##3}\VANthebibliography}

\usepackage{amsmath}
\usepackage{amssymb}
\usepackage{orcidlink}
\usepackage[nopatch]{microtype}
\usepackage{booktabs}
\usepackage{enumitem}
\usepackage{array} % text wrapping in tables
\usepackage{hyperref}
\hypersetup{
  colorlinks   = true, %Colours links instead of ugly boxes
  urlcolor     = blue, %Colour for external hyperlinks
  linkcolor    = blue, %Colour of internal links
  citecolor   = blue %Colour of citations
}
\usepackage[nameinlink,noabbrev]{cleveref}
\creflabelformat{equation}{#2\textup{(#1)}#3}
\crefname{figure}{Figure}{Figures}
\crefname{table}{Table}{Tables}

\title{Spectral signatures of young radio galaxies}

\author{Sophie A. Young~\orcidlink{0009-0000-5660-0302}}
\affiliation{School of Natural Sciences, Private Bag 37, University of Tasmania, Hobart, TAS 7001, Australia}
\email[S. A. Young]{sophie.young@utas.edu.au}
\author{Ross J. Turner~\orcidlink{0000-0002-4376-5455}}
\affiliation{School of Natural Sciences, Private Bag 37, University of Tasmania, Hobart, TAS 7001, Australia}
\author{Stanislav S. Shabala~\orcidlink{0000-0001-5064-0493}}
\affiliation{School of Natural Sciences, Private Bag 37, University of Tasmania, Hobart, TAS 7001, Australia}

\author{Georgia S. C. Stewart~\orcidlink{0000-0002-7155-6896}}
\affiliation{School of Natural Sciences, Private Bag 37, University of Tasmania, Hobart, TAS 7001, Australia}

\author{Patrick M. Yates-Jones~\orcidlink{0000-0003-2806-3495}}
\affiliation{School of Natural Sciences, Private Bag 37, University of Tasmania, Hobart, TAS 7001, Australia}

 %% First letter not capped

\begin{document}

\begin{abstract}
We investigate the evolution of active galactic nucleus jets on kiloparsec-scales due to their interaction with the clumpy interstellar medium (ISM) of the host galaxy and, subsequently, the surrounding circumgalactic environment. Hydrodynamic simulations of this jet--environment interaction are presented for a range of jet kinetic powers, peak densities of the multiphase ISM, and scale radii of the larger-scale environment -- characteristic of either a galaxy cluster or poor group. Synthetic radio images are generated by considering the combination of synchrotron radiation from the jet plasma and free-free absorption from the multiphase ISM. We find that jet propagation is slowed by interactions with a few very dense clouds in the host galaxy ISM, producing asymmetries in lobe length and brightness which persist to scales of tens of kpc for poor group environments. The classification of kiloparsec-scale jets is highly dependent on surface brightness sensitivity and resolution. Our simulations of young active sources can appear as restarted sources, showing double-double lobe morphology, high core prominence (CP~$> 0.1$), and the expected radio spectra for both the inner- and outer-lobe components. We qualitatively reproduce the observed inverse correlation between peak frequency and source size, and find that the peak frequency of the integrated radio spectrum depends on ISM density but not the jet power. Spectral turnover in resolved young radio sources therefore provides a new probe of the ISM. 
\end{abstract}

\section{Introduction}

Extragalactic radio jets launched from active galactic nuclei (AGNs) are among the most energetic phenomena in the Universe, interacting with their surroundings across many orders of magnitude in scale. These jets significantly impact the evolution of their host galaxies through kinetic and radiative feedback, acting both to suppress and enhance star formation \citep[][]{silk2013, wagner2016}. On larger scales, jets act to prevent catastrophic cooling in galaxy clusters \citep[e.g.,][]{mathews2003, croton2006, shabala2009, fabian2012, gaspari2012, schaye2015, raouf2017, weinberger2023}. The study of young radio sources and their evolution from compact to extended scales provides insight into how this feedback occurs, helping to constrain jet triggering mechanisms, radio source life cycles, and feedback channels.

Compact or unresolved sources make up a significant fraction of the observed radio source population, and are generally believed to form part of a similar evolutionary sequence to extended radio galaxies. However, many compact sources are likely to be frustrated by the dense gas located within their host galaxy, either temporarily or on longer timescales \citep[e.g.,][]{odea1997, orienti2016, odeasaikia2021}. Furthermore, studies of radio source lifetimes suggest that many young radio sources may not ever reach extended morphology. \citet{shabala2020} investigated empirical constraints on radio source duty cycles, finding that the observed fraction of remnant and restarted radio sources \citep[$\gtrsim 10\%$, e.g.,][]{jurlin2020} requires a population dominated by short-lived radio sources. This is consistent with the observed overabundance of compact, peaked-spectrum sources, which represent $15$--$30\%$ of the bright radio source population \citep[e.g.,][]{odea1997, orienti2016}. This fraction is a lower limit, since compact sources which dominate the number counts in the local Universe tend to have lower luminosities than their extended counterparts \citep{shabala2008,hardcastle2019}.

Gigahertz-peaked spectrum (GPS) and compact steep-spectrum (CSS) sources -- henceforth collectively referred to as peaked-spectrum (PS) sources -- are an important class of extragalactic radio source typically defined by their small linear sizes and turnover in their synchrotron radio spectra. GPS sources exhibit a peak at $\sim\!1$~GHz frequencies, with corresponding linear size $\mathrm{LS} \lesssim 1$~kpc. CSS sources are larger, with $1 \lesssim \mathrm{LS} \lesssim 20$~kpc, and often show a turnover or flattening of their radio spectra at lower ($\sim\!100$~MHz) frequencies \citep[see reviews by][]{odea1998, orienti2016, odeasaikia2021}. The small linear sizes associated with PS sources suggest that the jets interact with the dense, inhomogeneous gas located in the inner regions of the host galaxy. While the source is confined to these scales, the turnover in the radio spectrum is likely to be dominated by free-free absorption (FFA) due to the dense interstellar gas surrounding the radio source \citep[e.g.,][]{bicknell1997, callingham2015, tingay2015, bicknell2018}. Synchrotron self-absorption (SSA) is an alternative mechanism for the spectral turnover which constrains the internal properties of the source, such as the magnetic field or source size \citep[e.g.,][]{devries2009,jeyakumar2016,odeasaikia2021}. For the sources and environments investigated in this work, FFA is expected to be the more significant factor, and is the mechanism we consider in this work.

Models of FFA in the literature often assume a constant optical depth for the entire absorbing screen for simplicity, but these tend to be inconsistent with observations away from the spectral turnover \citep[e.g.,][]{readhead1996, odea1998, tingay2003, callingham2015}. Models which instead assume a power-law distribution of optical depths within the absorbing medium \citep[e.g.,][]{bicknell1997,kuncic1998} provide a much better fit to observational data. Conversely, many observers directly fit the spectra without explicitly assuming the underlying absorption mechanism or exact form of the absorbing medium \citep[e.g.,][]{snellen1998,callingham2017,kerrison2024}. 

Three-dimensional simulations of young radio sources interacting with an inhomogeneous multiphase ISM have allowed FFA spectra to be calculated directly from environment properties using realistic host galaxy gas distributions. \citet{bicknell2018} performed simulations of young radio sources expanding through the multiphase ISM, tracking the evolution of the turnover frequency assuming FFA as the absorption mechanism. Covering a range of jet and environment parameters, they found a relationship between turnover frequency and linear source size which is broadly consistent with the observed inverse correlation described by \citet{odea1997}.

Interactions with a clumpy medium also affect the morphology of the jets and the coupling of jet energy to the ISM and surrounding gas. A significant number of theoretical works have investigated the coevolution of AGN jets and their host galaxies through simulations, both for spherically symmetric clumpy gas distributions \citep[e.g.,][]{wagner2011,wagner2012,mukherjee2016,bicknell2018}, and in disk galaxies \citep[e.g.,][]{sutherland2007,gaibler2011,gaibler2012,mukherjee2018,tanner2022}. These small-scale interactions are important for star formation and jet feedback on galaxy scales, and are also likely to affect the eventual large-scale jet properties which are largely not investigated in the above simulation suites. With the exception of the simulations of \citet[][]{gaibler2011,gaibler2012} tracking jet evolution to linear sizes up to $\sim 30$~kpc, all of the above works only track jet evolution to several kpc, comparable to the size of the host galaxy.

By considering source expansion to linear sizes of up to $60$~kpc, in this paper we investigate the evolution of young radio sources to larger scales than probed in the literature, allowing simultaneous predictions regarding both the local emission and large-scale lobe structures. This paper builds on the results presented by \citet{bicknell2018} by considering how young radio sources evolve from compact to extended morphologies via interaction with their environments across a range of spatial scales, and by including both free-free absorption and semi-analytic synchrotron emissivity calculations based on the methods of \citet{turner2018} and \citet{yatesjones2022}. We robustly calculate synchrotron emission using Lagrangian tracer particles, and include synthetic observable quantities, such as surface brightness maps, with realistic resolutions comparable to currently available radio telescopes.

The paper is structured as follows. In Section~\ref{sec:methods}, we describe the simulation setup and emissivity calculations, then we present the simulations and results in Section~\ref{sec:results}. In Section~\ref{sec:asymmetry}, we consider the potential causes of lobe asymmetries and their persistence to larger scales for different circumgalactic environments, and discuss the effect of survey characteristics and the ambient environment on the classification of young radio sources in Section~\ref{sec:morphologies}. We investigate the evolution of peak frequency with source size for different jet and environment properties in Section~\ref{sec:seds}, and conclude in Section~\ref{sec:conclusion}. 

The spectral index $\alpha$ is defined by $S_\nu \propto \nu^{- \alpha}$ for flux density $S$ and frequency $\nu$ throughout this paper. We assume a $\Lambda$CDM concordance cosmology with $\Omega_{\rm M} = 0.307$, $\Omega_{\rm \Lambda} = 0.693$, and $H_0 = 67.7$~km~s$^{-1}$~Mpc$^{-1}$ \citep{planck2016}.

\section{Methods} \label{sec:methods}

In this section, we outline our numerical approach to capture the interactions between a jet and its host environment on both galactic and circumgalactic scales. First, this requires the development of a stable environment which is consistent with observations of both the multiphase ISM and larger-scale circumgalactic medium (CGM) surrounding young radio sources (Section~\ref{sec:env-setup}). Next, we discuss the process of injecting a jet into this environment and track its evolution by running numerical hydrodynamic simulations (Section~\ref{sec:sim-setup}). The observable properties of the source -- synchrotron surface brightness and radio spectra -- are then calculated by combining free-free absorption calculations with a semi-analytic calculation for the synchrotron emissivity (Section~\ref{sec:emis-calcs}).

\subsection{Environment setup}
\label{sec:env-setup} 

Most AGNs are hosted by elliptical galaxies \citep[][]{sadler1989,best2005}. While some previous work \citep[e.g.,][]{sutherland2007, gaibler2011, mukherjee2018} has focussed on the less ubiquitous case of jet--ISM interactions in disk galaxies, in this work, we follow the approach of \citet{bicknell2018}, and consider the host galaxy as a spherical region consisting of dense clouds embedded within the hot diffuse gas to represent the multiphase interstellar medium. The two phases are in pressure equilibrium, with the density of the clouds following a lognormal distribution \citep[as in][]{mukherjee2016}. The clouds are then placed under the influence of the same radial gravitational potential as the diffuse gas, and a density cutoff is imposed to remove clouds deemed to be thermally unstable (see Section~\ref{sec:dense-clouds} for our implementation). 

On large scales, most galaxy clusters are well-modelled by a $\beta$-profile \citep[e.g.,][]{king1962,cavaliere1976} to describe their density and gravitational potential. A number of models additionally include a `cusp', or other parameters beyond a simple $\beta$-profile, to more accurately represent the inner regions of the cluster \citep[e.g.,][]{zhao1996,nfw1996,vikhlinin2006}. In the following sections, we describe the implementation of a combined density profile for the diffuse gas, assuming a $\beta$-profile on circumgalactic scales (Section~\ref{sec:beta-profile}) and a double isothermal potential \citep[and associated density profile, e.g.,][]{sutherland2007} to describe the diffuse gas on host galaxy scales (Section~\ref{sec:double-isothermal}). This produces a smooth, spherically symmetric density profile for the diffuse gas across four orders of magnitude in spatial scale and provides the required modelling freedom in the central region. 

\subsubsection{Circumgalactic scales -- diffuse gas}\label{sec:beta-profile}

We model the density profile on circumgalactic scales using an isothermal $\beta$-profile \citep{king1962, cavaliere1976}, with associated gravitational potential as defined by \citet[][their equation 7]{krause2005}. The radial density profile is given by
\begin{equation}
    n_{\beta}(r) = n_{\beta, 0} \left( 1 + \left(\frac{r}{r_c}\right)^2 \right)^{-3\beta / 2},
\end{equation}
where $n_{\beta, 0}$ is the number density at radius $r = 0$, $r_c$ is the core radius of the galaxy cluster or group, and $\beta$ describes the steepness of the profile. An exponent of $\beta = 0.6$ and constant cluster temperature of $T_{\rm hot} = 10^7$~K were used throughout this work \citep[cf.][]{turner2015}.

To investigate the effect of large-scale environment on jet evolution and morphology, we consider core radii corresponding to both a galaxy cluster and a poor group. The core radius is related to the virial radius by $r_c = 0.08 r_{200}$ for a low-redshift cluster \citep[see][]{turner2015}, where $r_{200}$ defines the radius which encloses a mean density $200$ times greater than the critical value at that redshift. The mass within the virial radius is given by
\begin{equation}
    M_{200} = \frac{100}{G} H^2(z) r_{200}^3,
\end{equation}
where $G$ is the gravitational constant, and $H(z)$ is the Hubble constant at redshift $z$ \citep{croton2006}. Observed virial masses of typical galaxy clusters are $\sim \! 10^{14}$~M$_\odot$ \citep{vikhlinin2006}, corresponding to a virial radius of approximately $1$~Mpc for $z \sim 0$. Poor groups, meanwhile, have lower masses (and thus smaller virial and core radii), though these are generally less well-constrained, typically falling in the range $10^{11}$--$10^{13}$~M$_\odot$ \citep{tempel2014}. Using the ratio of core to virial radius defined by \cite{turner2015}, we choose core radii of $r_\mathrm{c, cluster} = 100$~kpc and $r_\mathrm{c, group} = 10$~kpc for the cluster and poor group respectively.

\begin{figure}
    \includegraphics[width=1.002\columnwidth,trim={6 7 5 4},clip]{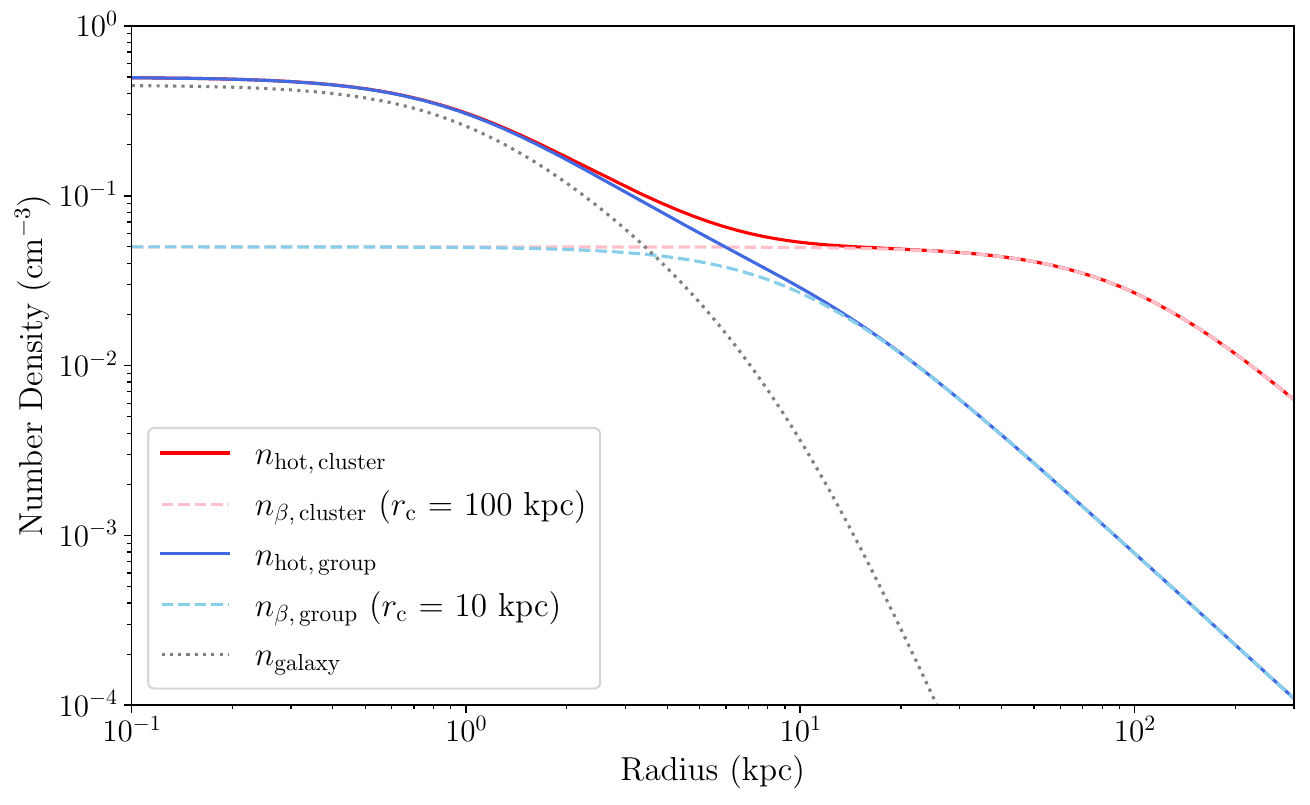}
    \caption{Density profiles for the diffuse gas in a galaxy cluster (red) and poor group (blue), each comprising an isothermal $\beta$-profile (dashed lines) and double isothermal galaxy profile (grey dotted).}
    \label{fig:cluster-group-density}
\end{figure}

On scales smaller than the core radius, the densities of the $\beta$-profiles are approximately constant (\cref{fig:cluster-group-density}, dashed lines). Using this density profile alone would thus not provide the modelling freedom required in the core region, particularly within the host galaxy ($r < 2.5$~kpc). Hence we introduce an additional profile for the host galaxy diffuse gas, as discussed in the following subsection.

\subsubsection{Host galaxy scales -- diffuse gas}\label{sec:double-isothermal}

Within the host galaxy, we model the hot diffuse gas of the multiphase ISM as spherically symmetric and isothermal. The density profile is defined using the potential derived by \cite{sutherland2007}, which combines the contributions from both baryonic and dark matter components. The net gravitational potential on host galaxy scales, $\phi_\mathrm{gal}(r)$, depends on parameters $\lambda = r_{\rm D}/r_{\rm B}$ and $\kappa = \sigma_{\rm D}/\sigma_{\rm B}$, where $r$ and $\sigma$ define the core radius and velocity dispersion, with subscripts `${\rm D}$' and `${\rm B}$' for dark and baryonic matter respectively. For normalised radius $r' = r/r_{\rm D}$ and potential $\psi = \phi/\sigma_{\rm D}^2$, the double isothermal potential is determined from Poisson's equation \citep[][their equation 5]{sutherland2007}:
\begin{equation}
    \label{eqn:poisson}
    \frac{\mathrm{d}^2 \psi}{\mathrm{d}r'^2} + \frac{2}{r'} \frac{\mathrm{d}\psi}{\mathrm{d}r'} = 9 \left[ \exp{(- \psi)} + \frac{\lambda^2}{\kappa^2} \exp{(-\kappa^2 \psi)} \right].
\end{equation}
This potential is used to determine the spherically symmetric density profile associated with the diffuse gas of the host galaxy ISM. In this work, we use values of $\lambda = 5$ and $\kappa = 2$ \citep[cf.][]{bicknell2018}. The resulting density profile is described by \citep[][their equation 2]{mukherjee2016}:
\begin{equation}
    n_\mathrm{gal}(r) = n_{\mathrm{gal},0} \exp{\left( -\frac{\mu m_{\rm a} \phi_\mathrm{gal}(r)}{k_{\rm B} T_\mathrm{hot}} \right) },
\end{equation}
where $n_{\mathrm{gal},0}$ is the density at the centre of the galaxy ($r=0$), $\mu=0.60364$ is the mean particle mass in atomic mass units ($m_\mathrm{a}$), $k_{\rm B}$ is Boltzmann's constant, and $T_\mathrm{hot} = 10^7$~K is the temperature of the hot gas.

The total hot gas density profile is then determined by combining this profile with the relevant cluster or group $\beta$-profile, with $n_{\rm hot}(r) = n_{\rm gal}(r) + n_\beta (r)$, and the corresponding gravitational potential given by $\phi(r) = \phi_{\rm gal}(r) + \phi_\beta (r)$. We choose a central hot gas density value of $n_0 = 0.5$~cm$^{-3}$, consistent with the range of values determined from X-ray observations of cluster halo gas densities \citep[typically falling in the range $0.01 < n_0 < 1$~cm$^{-3}$, e.g.,][]{allen2006, croston2008}.

\subsubsection{Host galaxy scales -- dense clouds} \label{sec:dense-clouds}

The warm clouds of the multiphase ISM are assumed to follow a lognormal density distribution, following the approach of \cite{mukherjee2016}. A lognormal fractal cube was generated using the freely available \textsc{pyfc}\footnote{\label{foot:pyfc}\url{https://pypi.org/project/pyFC/} (A. Wagner)} routine, with initial mean $\mu = 1$ and minimum and maximum correlated length scales of $20$~pc and $250$~pc respectively. Each location in the fractal cube has an associated density given by $n_{\rm fractal}(x, y, z)$, which is then scaled by a mean central density and placed under the influence of the same gravitational potential as the surrounding smooth gas. That is,
\begin{equation}
   n_\mathrm{warm}(x, y, z) = n_\mathrm{fractal}(x, y, z)\times n_{\mathrm{w}, 0} \exp{\left(- \frac{\phi(r)}{\sigma_{\rm t}^2} \right)},
\end{equation}
for spherically symmetric gravitational potential $\phi(r)$ and cloud velocity dispersion $\sigma_{\rm t}$. In this work, the `host galaxy' (i.e., the extent of the multiphase ISM) is taken to be a spherical region with a radius of $r_{\rm gal} = 2.5$~kpc.

We calculate the temperature at each location in the cloudy medium assuming that the warm clouds are initially in pressure equilibrium with the hot phase. This allows an upper limit of $T_\mathrm{crit} = 3.4 \times 10^4$~K to be placed on the temperature of any warm clouds; above this value, the clouds are taken to be thermally unstable and replaced with the corresponding hot gas values, as in \citealp{mukherjee2016}. Consistent with those authors, self-gravity of the dense clouds is not included in this work, however a combined $\beta$-profile and double-isothermal potential is implemented as described above to prevent atmospheric collapse. With our numerical setup (Section~\ref{sec:sim-setup}), these environments are stable for longer than the typical jet propagation time of several Myr. 

\begin{table}
    \centering
    \begin{tabular}{lll}
        \hline
        Parameter &  & Value \\
        \hline
        Halo temperature & $T_\mathrm{hot}$ & $10^7$~K \\
        Central halo density & $n_\mathrm{hot,0}$ & $0.5$~cm$^{-3}$ \\
        Central $\beta$-profile density & $n_{\beta,0}$ & $0.05$~cm$^{-3}$ \\
        $\beta$-profile slope & $\beta$ & $0.6$ \\
        Central galaxy-profile density & $n_\mathrm{gal,0}$ & $0.45$~cm$^{-3}$ \\
        Dark matter core radius & $r_{\rm D}$ & $5.0$~kpc \\
        Ratio of $r_{\rm D}/r_{\rm B}$ & $\lambda$ & $5$ \\
        Dark matter velocity dispersion & $\sigma_{\rm D}$ & $500$~km~s$^{-1}$ \\
        Ratio of $\sigma_{\rm D}/\sigma_{\rm B}$ & $\kappa$ & $2$ \\
        Spherical galaxy radius & $r_\mathrm{gal}$ & $2.5$~kpc \\
        Critical cloud temperature & $T_\mathrm{crit}$ & $3.4 \times 10^4$~K \\
        \hline
    \end{tabular}
    \caption{Environment parameters common to all simulations.}
    \label{tab:fixed-params}
\end{table}

\subsection{Simulation setup} \label{sec:sim-setup}

The simulations presented in this paper use the open-source numerical simulation package \textsc{pluto}\footnote{\label{foot:pluto}\url{http://plutocode.ph.unito.it/}} \citep{mignone2007}. We use the hydrodynamics (HD; non-relativistic) physics and Lagrangian particle modules of \textsc{pluto} version $4.3$, along with the HLLC Riemann solver, linear reconstruction, second-order Runge-Kutta time-stepping, the ideal equation of state, and the minmod limiter in the presence of shocks.

The simulations were performed on the \emph{kunanyi} high-performance computing facility provided by the Tasmanian Partnership for Advanced Computing. Each simulation used $2240$ cores, requiring an average of 100~000 CPU hours. The computational grid is defined by a three-dimensional Cartesian grid centred at the origin, extending to $\pm 35$~kpc in the $x$, $y$, and $z$ directions. The total number of grid cells in each direction is $n_{x,y,z} = 1000$, consisting of a central uniform grid patch of $500$ cells (from $-2.5 \rightarrow 2.5$~kpc) with a geometrically stretched grid extending for $250$ cells either side. The uniform grid patch has a resolution of $0.01$~kpc per cell, allowing both the jet injection region and dense clouds in the inner galaxy to be sufficiently resolved. The outer regions have a stretching ratio of $1.0159$, corresponding to a minimum resolution of $0.515$~kpc per cell at the outer boundary at $35$~kpc. Reflective boundary conditions are applied at all external boundaries. 

The jet injection region is defined as a sphere of radius $r_0 = 0.1$~kpc centred at the origin, following the geometry of the jet injection setup described by \citet{yatesjones2021,yatesjones2023}. Within this region, the pressure is matched to the ambient pressure at $r_0$, and the non-relativistic jet density is given by
\begin{equation} \label{eqn:jet-density}
    \rho_\mathrm{jet} = \frac{Q_\mathrm{jet}}{v_\mathrm{jet}^3 \pi r_0^2 \left(1 - \cos{\theta_\mathrm{jet}}\right) },
\end{equation}
where $Q_{\rm jet}$ is the one-sided jet power. For all simulations considered in this work, the jet speed is $v_\mathrm{jet} = 0.1c$, and the half-opening angle of the jet is $\theta_\mathrm{jet} = 10^\circ$. Each jet is injected onto the computational grid as a conical outflow, with the velocity within the injection region defined along the axis of jet propagation for both jets as $v = v_\mathrm{jet}$ if $\theta \leq \theta_\mathrm{jet}$, and $v = 0$ otherwise. An injection radius of $r_0 = 0.1$~kpc provides sufficient resolution to capture jet collimation dynamics. Lagrangian tracer particles -- required for our synchrotron emissivity calculations -- are uniformly distributed within the injection region, with $16$ particles injected onto the grid every $0.001$~Myr. 

Unlike some other works \citep[e.g.,][]{english2016,perucho2019,yatesjones2021}, we inject non-relativistic jets into the multiphase ISM. For the spatial scales and resolutions being considered in this work, mechanisms relevant to the stability of relativistic jets (such as helicity and magnetic fields) cannot be captured. The initially relativistic jets will rapidly decelerate and become mass-loaded at the edges via entrainment of the ambient gas \citep[][]{bicknell1984,bicknell1994}, forming a stable `spine-sheath' structure comprising a narrow, relativistic spine surrounded by a dense, slow-moving sheath. In our simulations we approximate this structure with a heavy, non-relativistic jet to robustly capture interactions with the multiphase ISM out to CGM scales. Following \citet{turner2023raise}, a typical jet with a mildly relativistic spine (e.g., Lorentz factor $\gamma = 2$, $v_{\rm spine} = 0.866 c$) will have an average bulk velocity of $\bar{v}_{\rm jet} = 0.1c$ across the cross-section of the jet (as modelled in our simulations). 

Cooling is not included in these simulations; we implicitly assume that unmodelled feedback channels such as stellar winds and radiative AGN feedback will act to prevent catastrophic cooling. This allows the heating-cooling balance to be maintained on large scales; however, the more complex dynamics on small scales are likely to be missed by this approach. Cooling of the dense gas clouds will partly mitigate ablation by the jet's passage \citep{antonuccio2008}, while jet dynamics will be affected by radiative shocks producing higher post-shock densities and slower shock propagation \citep{alexander2002,sutherland2017}. Inclusion of cooling in the simulation setup presented here is not feasible due to numerical instabilities and a significant increase in computational requirements; we defer further examination to future work.

\subsection{Emissivity calculations} \label{sec:emis-calcs}

We calculate synthetic surface brightness maps and radio spectra for each simulated source by combining synchrotron emissivities and tracer particle losses with free-free absorption calculations for a range of realistic observing parameters. Each simulation produces \textsc{pluto} grid and Lagrangian particle data files with temporal resolution of $0.05$ and $0.001$~Myr respectively. Grid data files contain the density, pressure, velocity, and jet tracer values for each cell on the \textsc{pluto} grid. Particle files contain the spatial coordinates, velocity, pressure, injection time, jet tracer value, and time since last shock, $t_{\rm acc}$, for each particle at a given timestep (see \citealt{yatesjones2022} for details). 

The synthetic lobe surface brightness is calculated using simulation grid and particle data, with each particle representing a packet of electrons whose properties evolve with the fluid. This approach calculates the lossless emissivity (including Doppler-boosting) based on the fluid variables in the simulation grid \citep[e.g.,][their equation 40]{turner2023raise} and weights these brightnesses by an adiabatic and radiative loss factor, $\mathcal{Y}(\boldsymbol{r}, t, t_{\rm acc}, \nu)$, derived from the Lagrangian particles \citep[cf.][]{turner2019}. That is, the lossless luminosity arising from a simulation grid cell at position $\boldsymbol{r}$ is given by
\begin{multline}
    \delta P_{\nu}(\boldsymbol{r}, t, t_{\rm acc}) = K(s) \nu^{(1-s)/2} \frac{q^{(s+1)/4}}{(q+1)^{(s+5)/4}}  \\
    \times p(\boldsymbol{r})^{(s+5)/4} \delta V(\boldsymbol{r}) \mathcal{Y}(\boldsymbol{r}, t, t_{\rm acc}, \nu),
\end{multline}
where $K(s)$ is a source-specific constant for electron energy injection index $s$ \citep{turner2018}, $q \equiv u_{\rm B}/u_e$ is the equipartition factor, $\delta V(\boldsymbol{r})$ is the volume of the grid cell, and $p(\boldsymbol{r})$ is the associated thermal pressure. The adiabatic and radiative (i.e., synchrotron radiation and inverse-Compton scattering) loss factor of the electron population is derived following \citet{yatesjones2022} and mapped to the simulation grid using a smoothing kernel with scale related to the local number density of particles \citep{stewart2024}. In this work, the jets are injected along the $z$-axis, with the observing plane chosen to be $xz$. Each pixel in the observing plane corresponds to a line of sight through the source (i.e., the integral of the grid cell luminosities along that line of sight).

To consider the effect of free-free absorption on the observed intensity, we must also calculate the optical depth associated with each line of sight through the observing grid. The optical depth is calculated following the approach of \cite{bicknell2018} using the \textsc{pluto} grid quantities density and pressure. The full expression for the emergent intensity along each line of sight is given by \citep[][their equation A2]{bicknell2018}:
\begin{equation}\label{eqn:intensity}
    I_\nu(s_2) = \int^{s_2}_{s_1} j_\nu(s') \exp{\left( -[\tau_\nu(s_2) - \tau_\nu(s')] \right)} \, \mathrm{d}s',
\end{equation}
where $s_1$ and $s_2$ are the locations of the back and front of the source respectively (with respect to the observer), $j_\nu(s)$ is the synchrotron emissivity associated with location $s$ at observing frequency $\nu$, and $\tau_\nu(s) = \int^s_{s_1} \alpha_\nu(s') \, \mathrm{d}s'$ is the free-free optical depth. The free-free absorption coefficient, $\alpha_\nu$, is a function of density, temperature, and observing frequency \citep[see][Appendix C]{bicknell2018}, calculated at every point on the observing grid using the \textsc{pluto} data outputs. That is,
\begin{equation}
    \label{eqn:abscoeff}
    \alpha_\nu(Z) = \sqrt{\frac{32 \pi}{27}} c^2 r_e^3 \left(\frac{k_\mathrm{B} T}{m_e c^2} \right)^{-3/2} n_e n_i(Z) Z^2 g_\nu(T,Z) \nu^{-2},
\end{equation}
where
\begin{equation}
    \label{eqn:gaunt}
    g_\nu(T,Z) = \frac{\sqrt{3}}{2 \pi} \left\{\ln{\left[ \frac{8}{\pi^2} \left(\frac{k_\mathrm{B} T}{m_e c^2} \right)^3 \frac{c^2}{r_e^2} \frac{1}{\nu^2 Z^2} \right]} - \sqrt{\gamma_E} \right\},
\end{equation}
for ions of charge $Ze$ where $Z$ is a positive integer (assumed to be $1$ for the remainder of this work). Here, $r_e$ is the classical electron radius, and $\gamma_E \approx 0.577$ is Euler's constant. We impose a temperature cutoff of $T = 1.05 \times 10^4$~K for FFA -- below this temperature, the gas is assumed to be neutral, and thus has its free-free absorption coefficient set to $\alpha = 0$. This choice of temperature cutoff is consistent with \citet{mukherjee2016}, and the rapid cooling expected below this temperature \citep{perucho2024}. The radio power at each observing frequency is calculated by integrating the emergent intensity over the two-dimensional observing grid.

\section{Results}\label{sec:results}

\begin{figure}
	\includegraphics[width=\columnwidth]{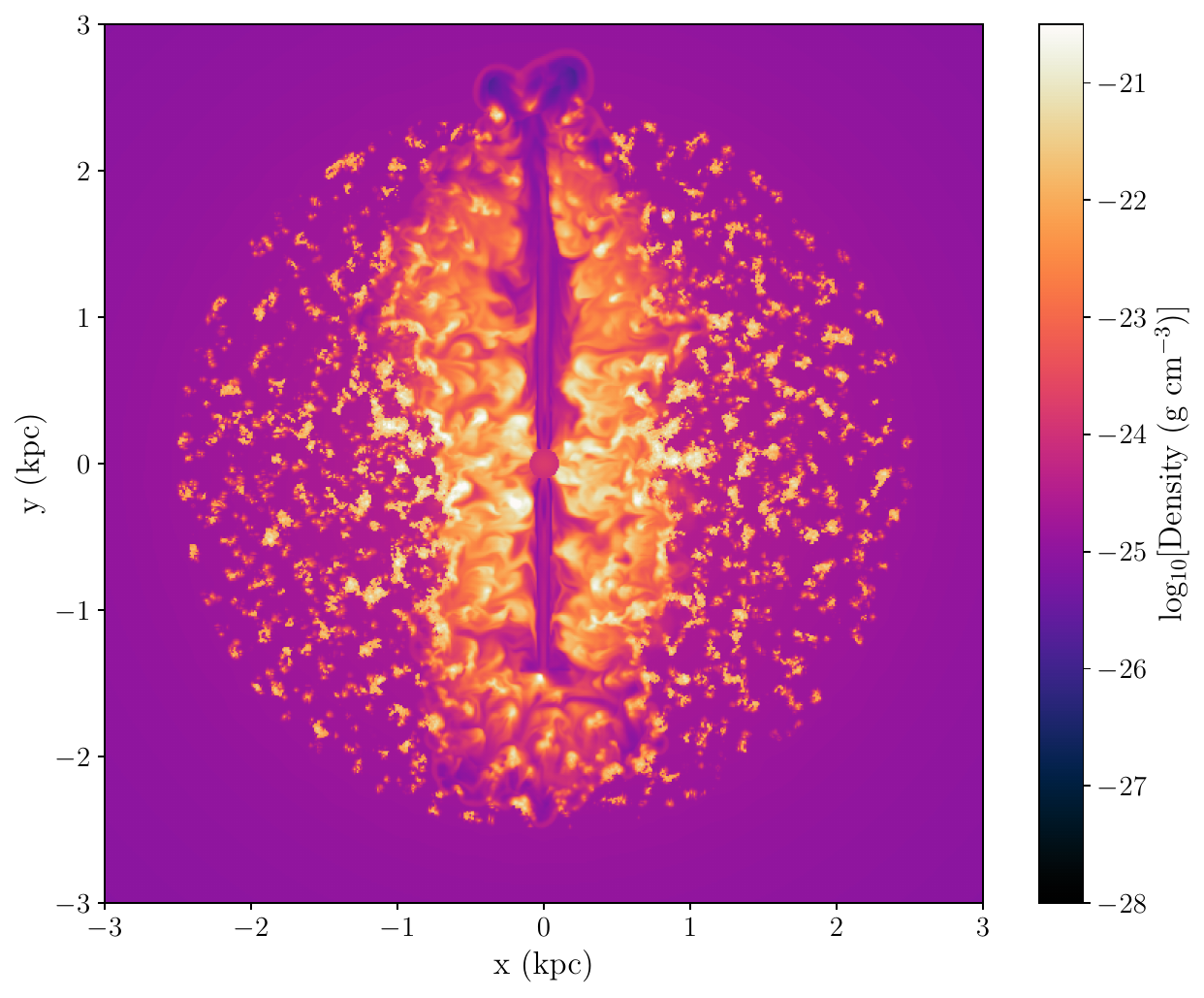}
    \caption{Midplane density slice for the `reference' simulation (n400-Q44) at a source age of $1$~Myr.}
    \label{fig:density-zoom}
\end{figure}

\begin{figure*}
	\includegraphics[width=\textwidth]{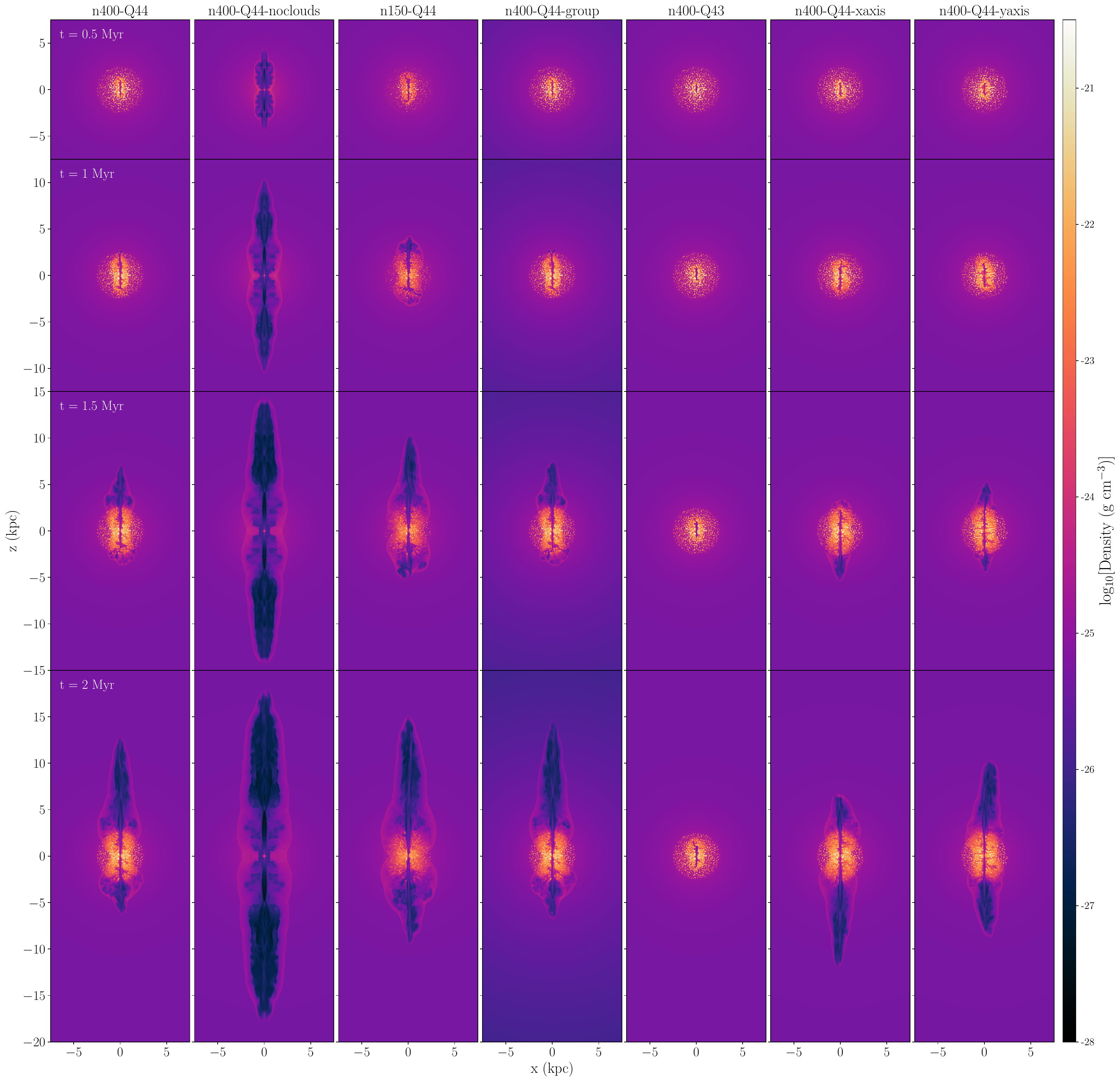}
    \caption{Midplane density slice for all simulations at source ages of $0.5$, $1$, $1.5$, and $2$~Myr.}
    \label{fig:density-grid}
\end{figure*}

\begin{figure*}
	\includegraphics[width=\textwidth]{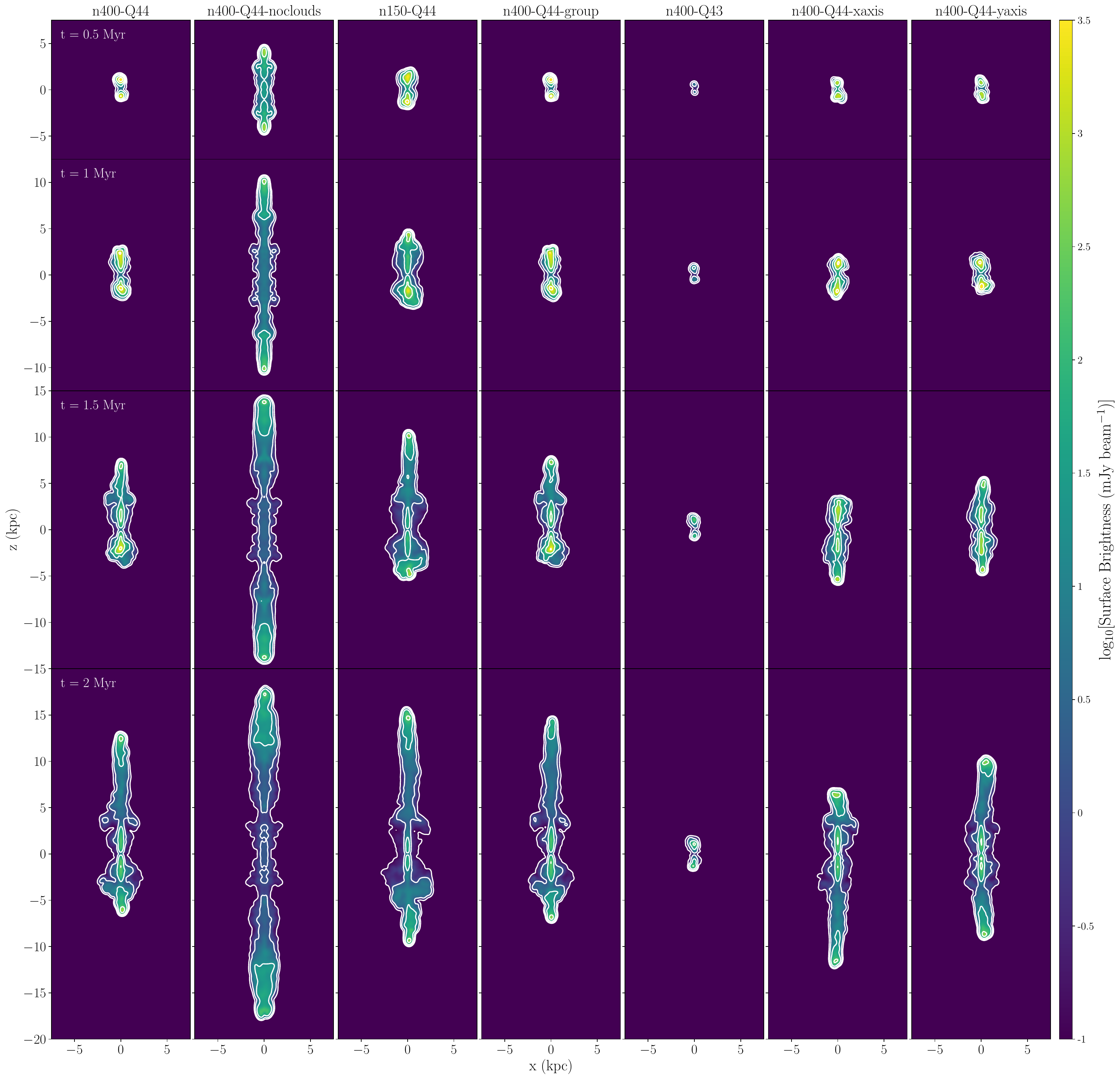}
    \caption{Synthetic radio emission for all simulations at source ages of $0.5$, $1$, $1.5$, and $2$~Myr. The observing frequency is $1.4$~GHz with a beam FWHM of $0.3$~arcsec; contours are spaced logarithmically by $1.1$~dex between $0.1$ and $3000$~mJy~beam$^{-1}$.}
    \label{fig:sb-grid}
\end{figure*}

In this section, we introduce the suite of simulations and their parameters, and present density and surface brightness maps for each of our simulated sources. The simulations used in this work are presented in \cref{tab:simulation_table}. Simulation labels of the form \textbf{nAAA-QBB-CCC} are used, where \textbf{AAA} gives the mean central density of the warm clouds in cm$^{-3}$ ($n_\mathrm{w, 0}$), \textbf{BB} defines the jet power in erg~s$^{-1}$ ($\log_{10}{Q_\mathrm{jet}}$), and \textbf{CCC} gives any additional information related to the surrounding environment. Each simulation is based on the properties of the `reference' simulation (n400-Q44, with a central cloud density of $400$~cm$^{-3}$, jet power of $10^{44}$~erg~s$^{-1}$, and large-scale cluster environment), with the change of one parameter. A zoom-in of the central region for n400-Q44 is presented in \cref{fig:density-zoom}. Midplane density slices and synthetic surface brightness maps for all simulations are presented in \cref{fig:density-grid,fig:sb-grid} respectively. 

\begin{table}
	\centering
	\caption{Simulation runs and their properties. Note: $r_{\rm c}=100$~kpc corresponds to a large-scale cluster environment, whilst $r_{\rm c} = 10$~kpc corresponds to a poor group environment.}
	\label{tab:simulation_table}
	\begin{tabular}{lccc>{\raggedright\arraybackslash}p{0.2\textwidth}}
		\hline
		 & $n_{\mathrm{w},0}$ & $Q_\mathrm{jet}$ & $r_\mathrm{c}$ & \\
        Sim. label & (cm$^{-3}$) & (erg~s$^{-1}$) & (kpc) & Type \\
		\hline
		n400-Q44 & $400$ & $10^{44}$ & $100$ & Reference simulation\\
		n400-Q44-noclouds & N/A & $10^{44}$ & $100$ & No clouds\\
		n150-Q44 & $150$ & $10^{44}$ & $100$ & Low density\\
        n400-Q44-group & $400$ & $10^{44}$ & $10$ & Group environment\\
        n400-Q43 & $400$ & $10^{43}$ & $100$ & Low jet power\\
        n400-Q44-xaxis & $400$ & $10^{44}$ & $100$ & Rotated environment\\
        n400-Q44-yaxis & $400$ & $10^{44}$ & $100$ & Rotated environment\\
		\hline
	\end{tabular}
\end{table}

Two `rotated environment' simulations (n400-Q44-xaxis and n400-Q44-yaxis) use the same jet and environment properties as n400-Q44, but with the host environment rotated by $90^\circ$. This is equivalent to injecting the jet along a different axis compared to the reference simulation and has the effect of changing the density profile encountered by the jet as it proceeds through the host galaxy. The effect of the rotated environment is discussed in detail in Section~\ref{sec:asymmetry-genesis}.

\cref{fig:density-grid} shows the evolution of density in the $xz$-plane ($y=0$) for each simulation. This highlights the morphological differences produced via the interaction with a dense host galaxy environment and the different growth rates. The radial symmetry in the environment of the n400-Q44-noclouds simulation produces a source which is symmetric throughout its lifetime, whilst large asymmetries between the jet and counterjet lengths are produced by the inhomogeneous dense clouds. We discuss this further in Section~\ref{sec:asymmetry}. 

Following the procedure outlined in Section~\ref{sec:emis-calcs}, simulation and particle outputs are converted to observable quantities, namely surface brightness maps and radio spectra. The evolution of the synthetic surface brightness for each simulation is shown in \cref{fig:sb-grid}, assuming an observing frequency of $1.4$~GHz, beam full-width at half-maximum (FWHM) of $0.3$~arcseconds, and a redshift of $z = 0.05$. These parameters are representative of the \emph{enhanced Multi-Element Radio Linked Interferometer Network} (eMERLIN), assuming a sensitivity similar to that of the Legacy eMERLIN Multi-band Imaging of Nearby Galaxies survey \citep[LeMMINGs;][]{lemmings2018}. At later times, there appears to be a significant contribution to the emission from the spatial region encompassing the host galaxy, as well as lobe structures on larger scales. This contribution is heightened for simulations which contain a dense multiphase ISM; we discuss this effect in Section~\ref{sec:core-lobe}.

\begin{figure}
	\includegraphics[width=1.002\columnwidth,trim={6 7 5 4},clip]{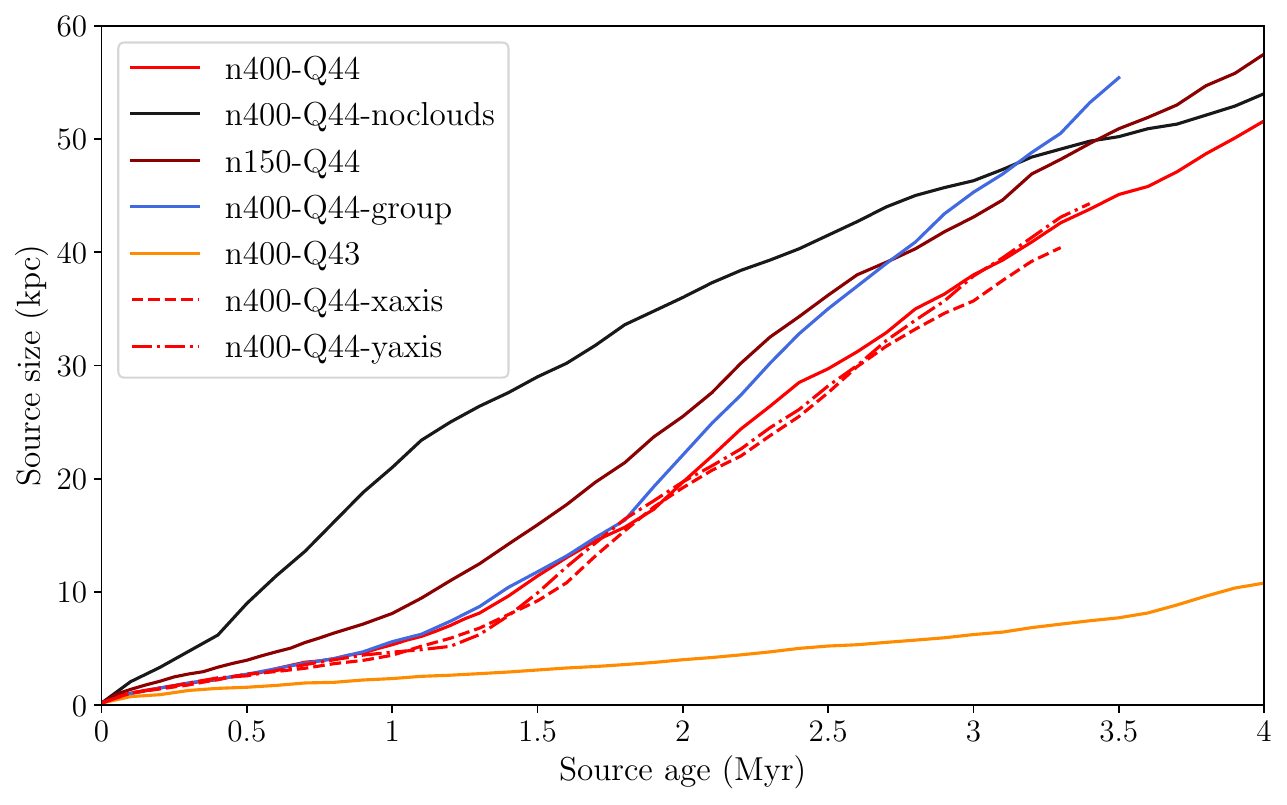}
    \caption{Total source size as a function of source age for all simulations.}
    \label{fig:length-time}
\end{figure}

\begin{figure}
	\includegraphics[width=1.002\columnwidth,trim={6 7 5 4},clip]{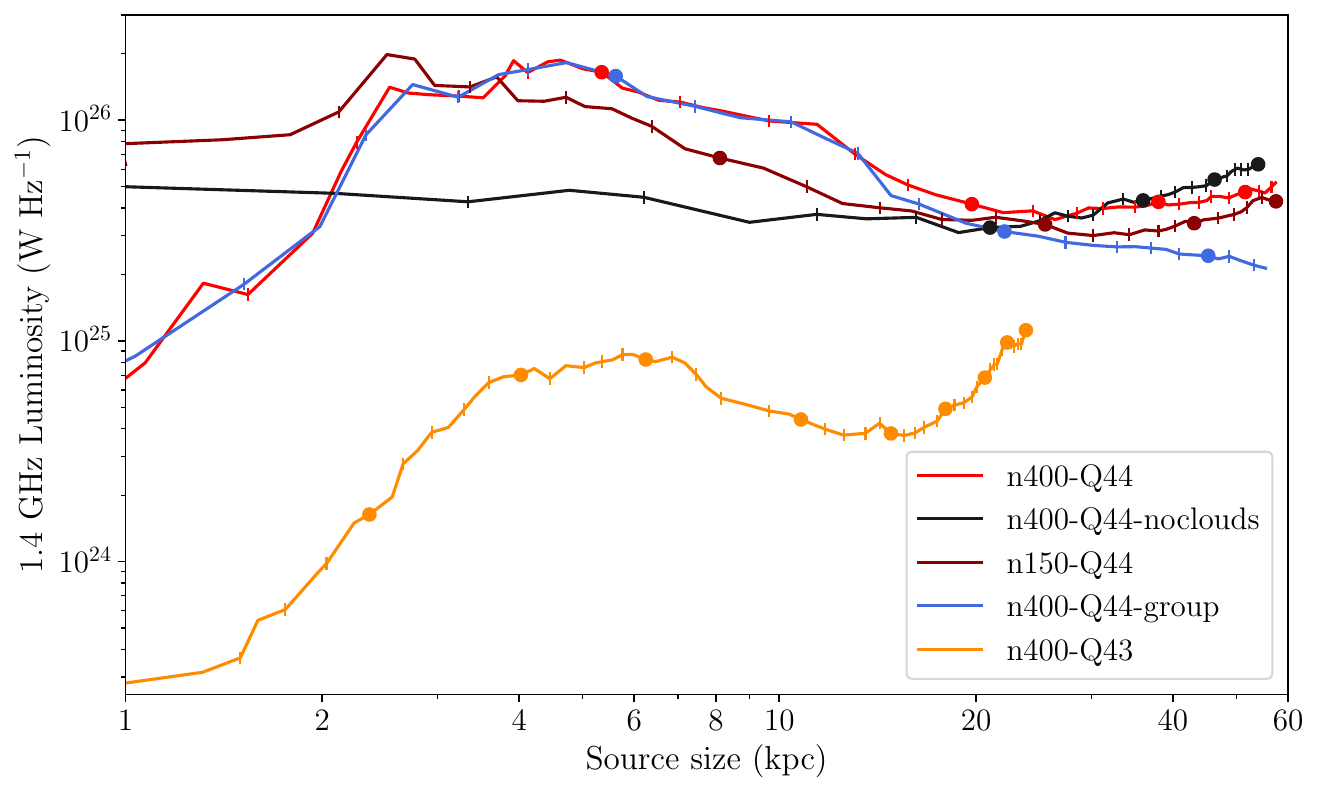}
    \caption{Total source luminosity at $1.4$~GHz as a function of source size for all simulations. Rotated environment simulations (n400-Q44-xaxis and n400-Q44-yaxis) are not included for clarity; these show very similar evolution to n400-Q44. Vertical markers are shown every $0.2$~Myr, with filled circles every $1$~Myr.}
    \label{fig:luminosity-size}
\end{figure}

In \cref{fig:length-time}, we show the evolution in the length of the radio source as a function of source age. Assuming infinite surface brightness sensitivity, the size of each source is taken to be the distance between the furthest extents of jet material in each direction. In Section~\ref{sec:asymmetry}, we quantify the asymmetry of each source via the jet to counterjet length ratio, and discuss the formation and persistence of asymmetries in both jet length and lobe brightness. We present the evolution of the source luminosities at $1.4$~GHz in \cref{fig:luminosity-size}. The effects of environment and jet properties on source growth rate can be summarised as follows: 

\begin{enumerate}[label=(\roman*)]
    \item A source expanding through a dense, inhomogeneous ISM experiences much slower growth than one expanding through a smooth medium (i.e., n400-Q44-noclouds). A higher host galaxy ISM density further decreases source growth rates (i.e., the jets in the n400-Q44 simulation are shorter at a given time than jets in the n150-Q44 simulation).
    \item At large source sizes, the three simulations with different ISM densities produce sources which are morphologically similar. Interaction with a dense ISM, however, results in a source being older for the same source size. Because jet power scales with the mass injection rate into the jet for a fixed jet velocity, older sources at the same size have higher luminosity. 
    \item For the same host galaxy environment, n400-Q44 and n400-Q44-group show identical growth within the host galaxy. Once both sources have reached the edge of the host galaxy ($t \approx 1.5$~Myr), the steeper density profile of the large-scale poor group environment results in faster source growth at later times.
    \item A low power jet grows more slowly through the same environment (i.e., the n400-Q43 jets are smaller at a given age than the n400-Q44 jets). This is due to both the lower rate of energy injection and the greater density contrast between the jet and environment material (since jet density is directly proportional to the jet power for a given jet speed; see \cref{eqn:jet-density}).
    \item The effect of rotating the environment on source growth rate is small, but present due to some jets interacting more strongly with the clouds in their path (i.e., n400-Q44 vs n400-Q44-xaxis vs n400-Q44-yaxis).
\end{enumerate}

\section{Lobe asymmetries} \label{sec:asymmetry}

Asymmetries in the length and brightness of the jets and lobes occur due to interactions with the dense cloud structures in the inner regions of the galaxy (see Section~\ref{sec:results}). In the following sections, we discuss the formation of this asymmetry as a result of jet--ISM interactions within the host galaxy (Section~\ref{sec:asymmetry-genesis}) and its persistence on larger scales (Section~\ref{sec:asymmetry-persistence}).

\subsection{Genesis of asymmetry}\label{sec:asymmetry-genesis}

\subsubsection{Host galaxy environment asymmetry}

Within the host galaxy, the jet and counterjet interact with dense clouds in their path, particularly along the jet axis. Since the distribution of dense clouds is not radially symmetric, differences between the environments encountered by the jet and counterjet result in asymmetries in both length and brightness.

\begin{table*}
    \centering
    \begin{tabular}{lcccccc}
        \hline
         & $m_+$ & $m_-$ &  & $t_+$ & $t_-$ &  \\
        Sim. & ($\times 10^{8}$~M$_\odot$) & ($\times 10^{8}$~M$_\odot$) & Mass ratio & (Myr) & (Myr) & Time ratio \\
        \hline
        n400-Q44 & $0.7332$ & $1.0628$ & $0.6898$ & $0.95$ & $1.60$ & $0.5937$ \\
        n400-Q44-xaxis & $0.9053$ & $0.8677$ & $1.0433$ & $1.45$ & $1.10$ & $1.3182$ \\
        n400-Q44-yaxis & $1.0017$ & $1.2173$ & $0.8229$ & $1.20$ & $1.30$ & $0.9231$ \\
        \hline
    \end{tabular}
    \caption{Environment asymmetry and jet/counterjet propagation times for simulations n400-Q44, n400-Q44-xaxis, and n400-Q44-yaxis. $m_+$ and $m_-$ refer to the total mass located within a cylinder of radius $0.1$~kpc with $|z| \leq 2.5$~kpc for $z > 0$ and $z < 0$ respectively. Mass ratio is defined by MR $= m_+/m_-$. $t_+$ gives the time for the northern jet to reach $z = 2.5$~kpc, $t_-$ gives the time for the counterjet to reach $z = - 2.5$~kpc. Time ratio is given by TR $= t_+/t_-$.}
    \label{tab:mass_values}
\end{table*}

We quantify this asymmetry in the environment by considering the mass located within a cylinder aligned with the jet axis for each of the reference simulations (i.e., those with $n_{\mathrm{w},0} = 400$~cm$^{-3}$, $Q_{\rm jet} = 10^{44}$~erg~s$^{-1}$, and a large-scale cluster environment: n400-Q44, n400-Q44-xaxis, and n400-Q44-yaxis). We choose a cylindrical radius of $0.1$~kpc for mass calculations, as this is approximately equal to the collimated jet radius. The masses encountered by the jet and counterjet in each simulation are shown in \cref{tab:mass_values}, along with the time taken for each of the jet heads to reach the edge of the host galaxy ($z = \pm 2.5$~kpc). The time for which the jet is confined depends primarily on the distribution of dense material encountered. Notably, despite being the most symmetric in terms of total mass along the jet axis, n400-Q44-xaxis has a significant time delay between the jet and counterjet reaching the edge of the host galaxy. This delay is due to the exact distribution of the dense gas along the path of the jet, and results in a source which is highly asymmetric (\cref{fig:density-grid}). Despite showing very similar growth rates in the total source size (\cref{fig:length-time}), the three rotated environment simulations exhibit very different behaviours in terms of their observed asymmetries on $10$--$30$~kpc scales; we discuss this further in Section~\ref{sec:asymmetry-persistence}.

\subsubsection{Jet--cloud interactions}

The interactions between the jet and clouds (and thus the resulting length asymmetries) also depend on the nature of the jet. For the reference environment ($n_{\rm w, 0} = 400$~cm$^{-3}$), this effect is investigated for the two jet powers (simulations n400-Q44 and n400-Q43). In both cases, we find that each jet is held up by one particularly dense region in its path. For the primary jet ($z > 0$)\footnote{The primary jet (i.e., the longer of the two jets on intermediate scales) is the northern jet in all of our simulations except n400-Q44-xaxis.}, this `northern cloud' is centred at $(x, y, z) = (-0.015, 0.005, 1.075)$~kpc, with an initial mass of $3.0 \times 10^7$~M$_\odot$ and maximum density of $5.0 \times 10^{-21}$~g~cm$^{-3}$. For the counterjet, the `southern cloud' is centred at $(x, y, z) = (-0.025, 0.015, -1.445)$~kpc, with an initial mass of $2.5 \times 10^7$~M$_\odot$ and peak density of $1.8 \times 10^{-20}$~g~cm$^{-3}$. Midplane density slices for each dense cloud are shown in \cref{fig:cloud-slices-combined}.

\begin{figure}
	\includegraphics[width=1.002\columnwidth,trim={6 5 5 4},clip]{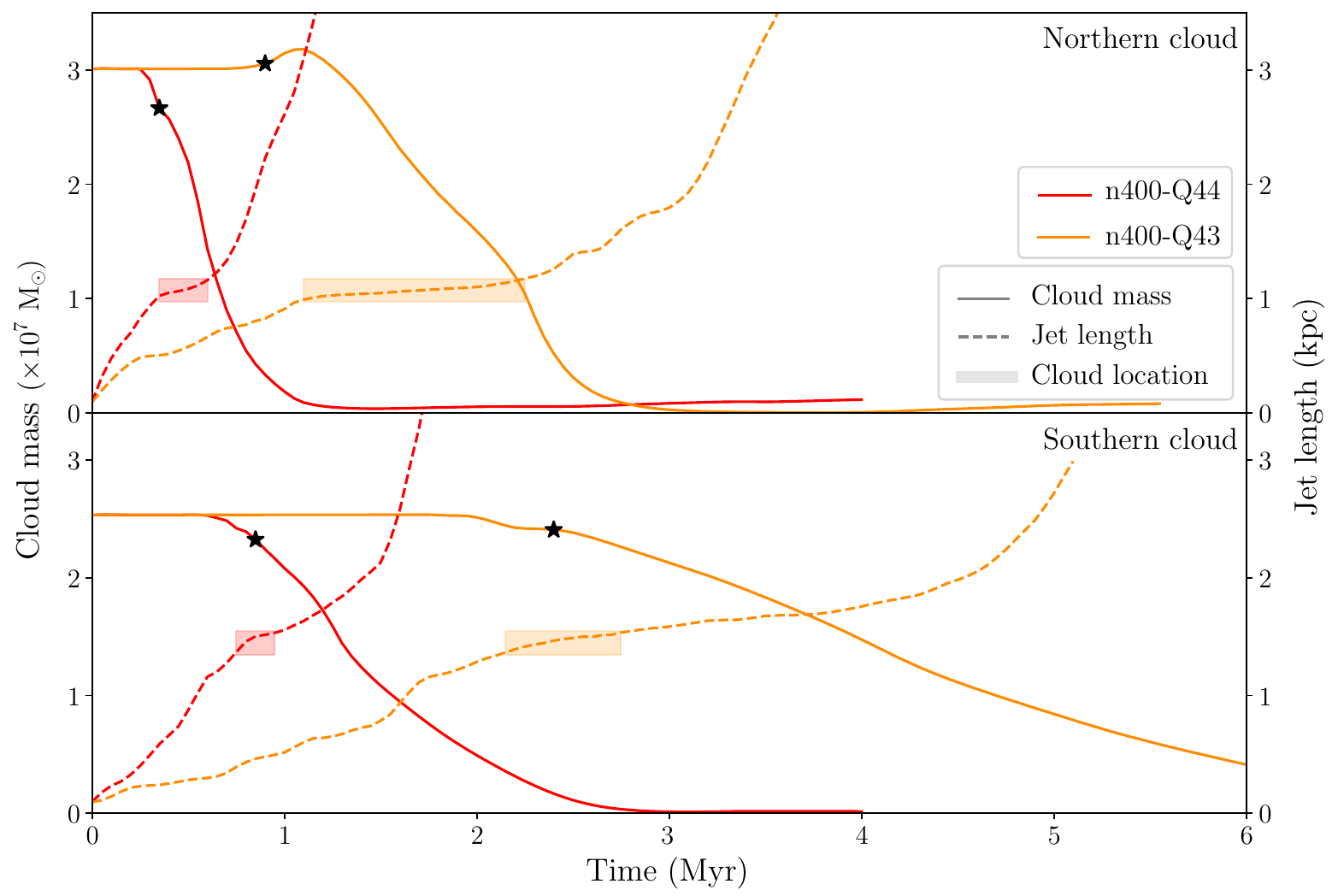}
    \caption{Evolution of total mass of cloud material (solid, left axes) and jet length (dashed, right axes) for northern and southern dense clouds. The black stars are the first timestep where the thermal pressure of cloud material exceeds the ram pressure of jet. The shaded regions correspond to the times where the jet is co-located with the dense cloud (within $\pm 0.1$~kpc).}
    \label{fig:cloud-mass}
\end{figure}

The main factors which affect individual jet--cloud interactions are the location and spatial extent of the cloud, its peak density and total mass, and the properties of the jet. For the two clouds discussed above, the southern cloud is smaller and slightly offset from the jet axis, so momentarily slows down and deflects the jet, whereas the larger northern cloud occupies effectively the entire jet channel, preventing the jet from progressing without destroying it entirely (see~\ref{app:extra-figures}). This can be understood by considering the time for which the jet head is at the exact location of the dense cloud in each case (i.e., shaded regions in \cref{fig:cloud-mass}). For the northern cloud, the jet velocity is reduced to effectively zero over a longer time-scale than for the southern cloud. We note, however, that the northern jet velocity (i.e., slope of the dashed line) increases rapidly after the jet passes the cloud. Conversely, the southern jet is slowed to a small but non-zero velocity due to the deflection of the jet around the cloud, eventually increasing once the cloud has been sufficiently ablated by the jet. The travel time of the jet over a small distance ${\rm d}s$ can be described by \citep[cf.][equation 5]{turner2023raise}
\begin{equation} \label{eqn:timescale}
    v = \frac{{\rm d}s}{{\rm d}t} = {v_{\rm jet}} \Bigg(\;\!1 + \bigg[ \frac{\bar{\rho}_x}{\gamma_{\rm jet}^2 \rho_{\rm jet}} \bigg]^{1/2\,} \Bigg)^{-1},
\end{equation}
where $v_{\rm jet}$ is the bulk velocity of the jet, $\bar{\rho}_x$ is the average density encountered by the jet-head, $\gamma_{\rm jet}$ is the Lorentz factor ($\approx 1$ for a non-relativistic jet), and $\rho_{\rm jet}$ is the jet density. Since the collision between the jet and the northern cloud is head-on, this implies a higher average density \emph{within} the jet channel, resulting in a lower forward velocity $v$ during the interaction. 

\cref{fig:cloud-mass} shows the evolution of the cloud mass for the north and south dense clouds. For both simulations considered here, the northern cloud loses mass faster than the southern cloud, due to the fact that it is completely ablated by the jet during the head-on interaction. Conversely, the southern cloud slows down the jet for longer (this is particularly evident in the case of the low jet power simulation, n400-Q43), resulting in a significant time delay between the jet and counterjet escaping the dense host galaxy. As discussed in the previous section, this produces noticeable length asymmetries which persist to larger scales.

For both clouds, a jet power of $10^{43}$~erg~s$^{-1}$ results in a slower mass loss than its high-power counterpart. This jet has a lower ram pressure ($p_{\rm ram} = \rho_{\rm jet} v_{\rm jet}^2$) and a higher density contrast between the jet and cloud material, since jet density is proportional to jet power (see \cref{eqn:jet-density}). This also results in a lower jet-head advance speed and longer interaction time (\cref{eqn:timescale}). For both jet powers, the thermal pressure of each cloud is enhanced during the jet--cloud interaction. This enhancement occurs when the advancing high pressure shock-front reaches the cloud and persists until the cloud has been completely ablated by the jet.

\subsection{Persistence}\label{sec:asymmetry-persistence}

\begin{figure}
	\includegraphics[width=1.002\columnwidth,trim={6 7 5 4},clip]{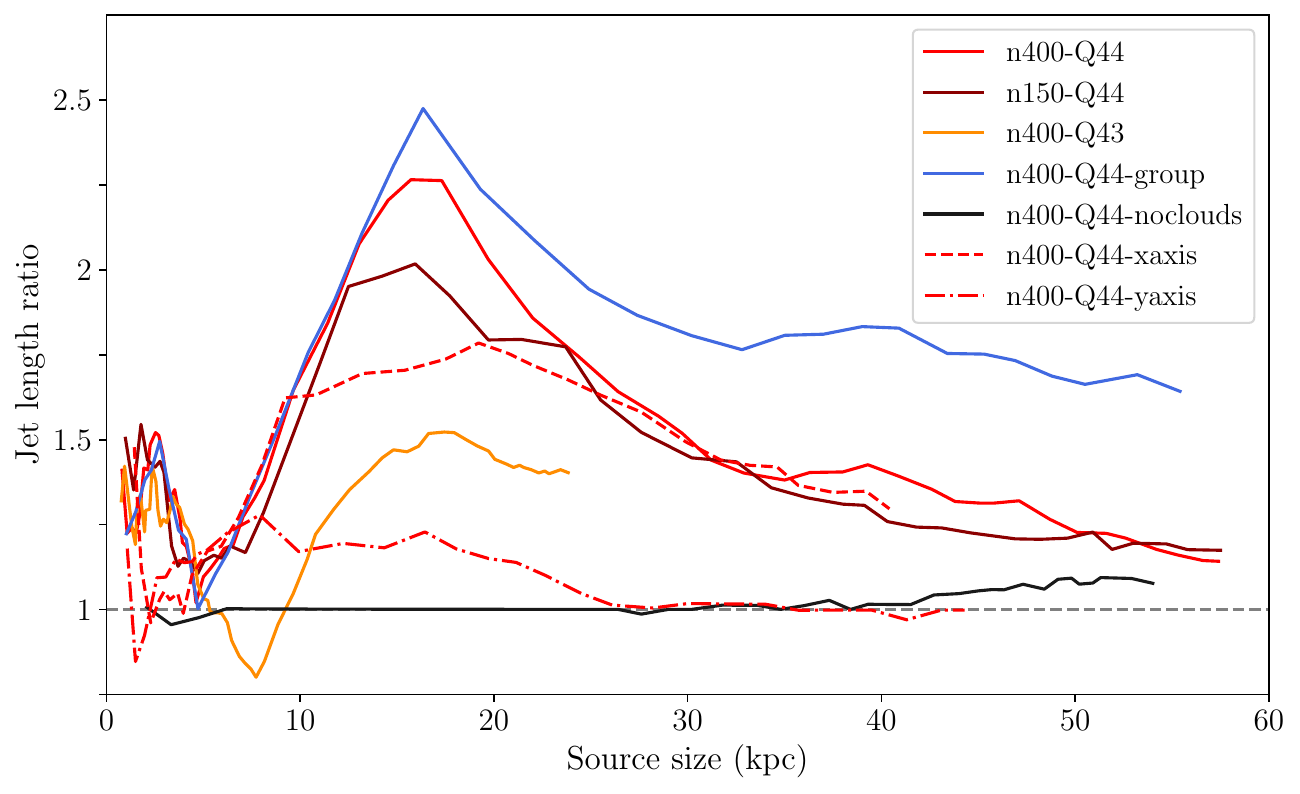}
    \caption{Length asymmetry between jet and counterjet, defined as $l$(primary jet)/$l$(counterjet) for jet length $l$. We note that the primary jet is always the longer of the two jets; this is the southern jet for simulation n400-Q44-xaxis, and the northern jet for the remaining simulations.}
    \label{fig:length-asymmetry}
\end{figure}

On larger scales, length and brightness asymmetries formed within the host galaxy remain present for some time after the jets expand past the edge of the host galaxy. As discussed in Section~\ref{sec:asymmetry-genesis}, the direction and magnitude of the asymmetry depend on the delay between the jet and counterjet escaping the host galaxy. \cref{fig:length-asymmetry} shows the jet length ratio (LR) as a function of total source size, defined as the length of the primary jet divided by that of the counterjet for all simulations.\footnote{We note that for simulation n400-Q44-xaxis, this length ratio is defined as the length of the southern jet divided by the northern jet length, so that LR~$> 1$ outside of the host galaxy.} In the case of n400-Q44-noclouds, the symmetry of the environment on both small and large scales results in a source which is symmetric throughout its active lifetime.

For sources expanding into a large-scale cluster environment (all but n400-Q44-group), the length ratio peaks at a source size of $15$--$20$~kpc before decreasing to approximately unity as the source continues to expand. For the source expanding into a large-scale group environment, the evolution to the peak LR is very similar to that of n400-Q44 with the same host galaxy environment, but this persists with the primary jet being approximately double the length of the counterjet for the remainder of the simulated lifetime of the source. 

The persistence of a significant length asymmetry can be understood by comparing the density profiles of the two environments (see \cref{fig:cluster-group-density}). The cluster environment has a core radius of $r_\mathrm{c} = 100$~kpc; for the group environment, this is only $10$~kpc. The density profile is effectively flat between $10$ and $100$~kpc for the cluster environment, allowing the shorter jet to `catch up' to the longer one. Conversely, in the group environment the jet quickly passes the flat inner region of the $\beta$-profile, so the longer jet is expanding into an environment with rapidly decreasing density in this region. 

We compare this result to the analytical model of \cite{kaiseralexander1997}, who showed that for a power-law density profile of the form $\rho(r) \propto r^{-\beta}$, the lobe expansion velocity varies with time as $v \propto t^{(\beta - 2)/(5 - \beta)}$. Immediately upon escaping the dense host galaxy medium ($2.5 \lesssim r \lesssim 10$~kpc; see \cref{fig:cluster-group-density}), the density profile in our simulations scales approximately as $\beta = 2$, implying a constant ballistic expansion velocity. Expansion through the cluster environment on $10$--$100$~kpc scales corresponds to $\beta \approx 0$, implying a decreasing velocity of $v \propto t^{-2/5}$. Meanwhile, the group environment in this region has a steeper density profile ($\beta \approx 1$), resulting in a lobe expansion velocity which is closer to constant, so the morphologies present immediately outside the host galaxy are maintained to much larger spatial scales.

\begin{figure}[t]
	\includegraphics[width=1.002\columnwidth,trim={6 7 5 4},clip]{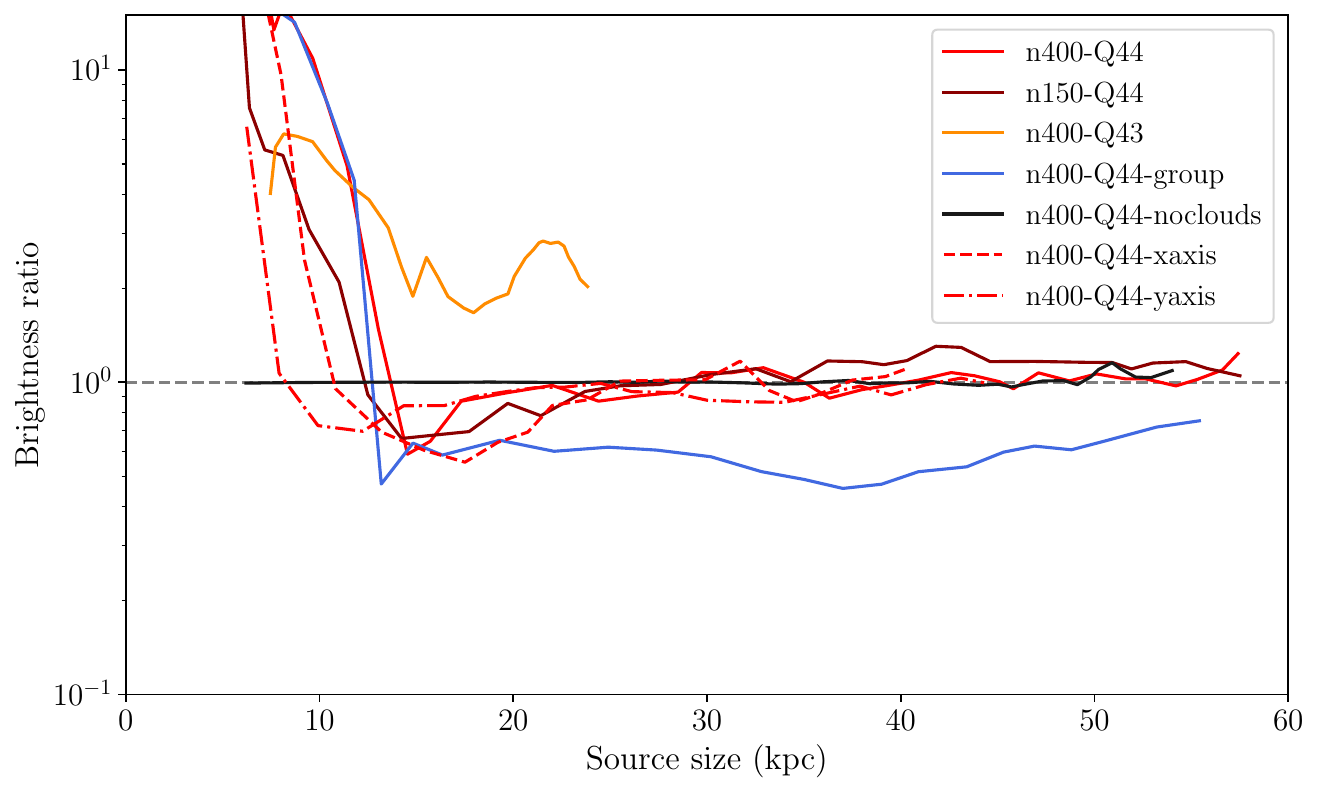}
    \caption{Lobe brightness asymmetry, defined as $P_{1.4~\mathrm{GHz}}$(primary lobe)/$P_{1.4~\mathrm{GHz}}$(secondary lobe) for $r > 2.5$~kpc. As in \cref{fig:length-asymmetry}, the primary lobe has $z < 0$ for simulation n400-Q44-xaxis.}
    \label{fig:brightness-asymmetry}
\end{figure}

\cref{fig:brightness-asymmetry} shows the lobe brightness ratio (BR) for each simulation as a function of total source size. Excluding emission from within the host galaxy (i.e. $r < 2.5$~kpc), this is defined as the $1.4$~GHz radio power of the primary lobe divided by that of the secondary lobe. As with the length ratio, all simulations (except n400-Q43) show a peak in BR at a source size of approximately $15$~kpc. Taking each ratio as the length or brightness of the primary jet divided by that of the secondary jet, this results in a peak LR greater than $1$, whilst the peak BR is generally less than $1$. That is to say, the shorter jet (at a source size of $\sim\!15$~kpc) tends to have brighter associated lobe emission, due to interaction with a more dense external environment. This is consistent with the results of hydrodynamic simulations of powerful radio sources expanding into asymmetric large-scale environments \citep{yatesjones2021} and observed trends for asymmetric sources studied by the Radio Galaxy Zoo project \citep{rodman2019}.

Our results suggest that extended sources with large observed length or lobe brightness asymmetries are more likely to be located in a poor group environment. While asymmetries in these simulations are developed \emph{within} the host galaxy, the flat density profile observed in galaxy clusters with large core radii acts to return length and brightness ratios to approximately unity within only a few kpc of reaching their peak values. Conversely, despite showing similar evolution to source sizes of $\sim 20$~kpc, a poor group environment acts to maintain asymmetries on larger scales.

\section{Interpretation of complex radio morphologies} \label{sec:morphologies}

\begin{figure*}
    \includegraphics[width=\textwidth]{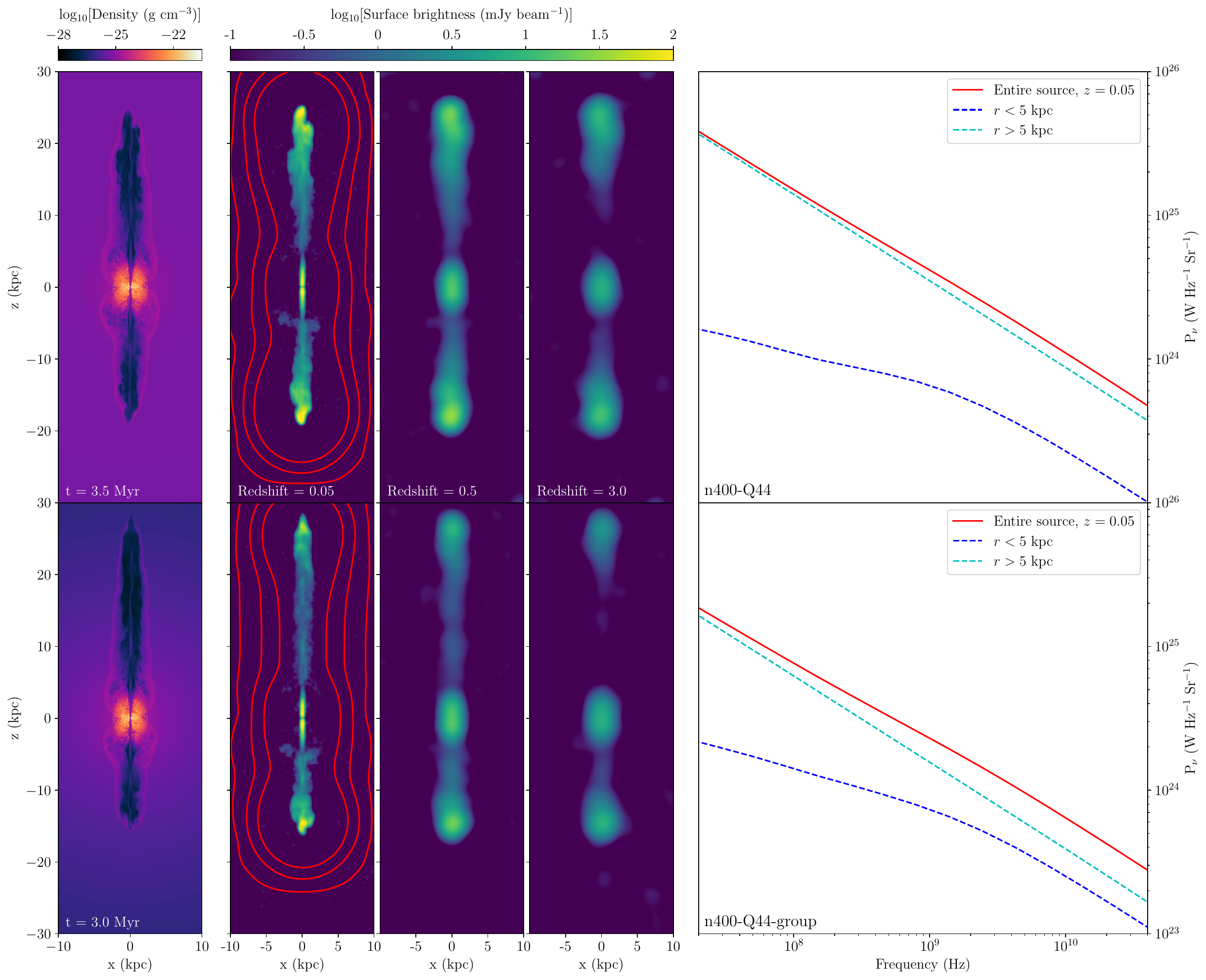}
    \caption{First column: Midplane density slice for simulations n400-Q44 (at $t = 3.5$~Myr, top row) and n400-Q44-group (at $t = 3.0$~Myr, bottom row). Second column: Synthetic surface brightness at a redshift of $z = 0.05$. Red contours represent emission detected by LOFAR at $150$~MHz, with a beam FWHM of $5$~arcsec (corresponding to $5$~kpc at this redshift). Contours are at $0.1$ \citep[cf. $5\sigma$ sensitivity from LoTSS;][]{shimwell2017}, $10$, and $100$~mJy~beam$^{-1}$. The colourbar represents emission detected by eMERLIN at an observing frequency of $1.4$~GHz. A beam FWHM of $0.3$~arcsec is assumed, and Gaussian noise has been added to the image with rms~$= 70$~$\mu$Jy~beam$^{-1}$ \citep[cf. median rms from LeMMINGs;][]{lemmings2018}. Third and fourth columns: Synthetic surface brightness detected by eMERLIN at redshifts of $z = 0.5$ and $z = 3$, respectively. LOFAR contours are not included as the beam FWHM corresponds to approximately $31$ and $40$~kpc at these redshifts, and hence the sources are unresolved. Fifth column: Synchrotron radio spectra for the entire source at a redshift of $z = 0.05$ (red), the inner lobe pair ($r < 5$~kpc, blue), and the outer lobe pair ($r > 5$~kpc, cyan). Spectra are calculated assuming infinite sensitivity at all frequencies.}
    \label{fig:ddrg-plot}
\end{figure*}

Surface brightness maps and radio spectra may not always be straightforward to interpret. In the following sections, we discuss the classification of radio galaxies based on observed morphology and how this is affected by survey resolution and surface brightness sensitivity (Section~\ref{sec:sensitivity}), as well as high core prominence values which are typically attributed to multiple bursts of jet activity (Section~\ref{sec:core-lobe}).

\subsection{Observational effects} \label{sec:sensitivity}

The correct classification of extended radio galaxies with prominent core emission may be confounded by poor resolution or surface brightness sensitivity. Below, we investigate the potential for such objects to be misclassified.

The midplane density slice, synthetic surface brightness, and synchrotron radio spectrum for the reference simulation (n400-Q44, $t = 3.5$~Myr) and the group environment simulation (n400-Q44-group, $t = 3.0$~Myr) are shown in \cref{fig:ddrg-plot}. We calculate the surface brightness assuming both LOFAR- and eMERLIN-like observations ($5$~arcsec and $0.3$~arcsec beam sizes respectively, and a sensitivity of $0.1$~mJy~beam$^{-1}$; see figure caption for details) at redshifts of $0.05$, $0.5$, and $3.0$. When observed with LOFAR at $150$~MHz, both sources appear only marginally resolved at $z = 0.05$, and unresolved at the higher redshifts, making it difficult to to draw conclusions about the nature of the sources from these observations alone. Follow-up observations with a high-resolution array, such as eMERLIN, can provide more detailed information about the structure of the source, however these instruments typically have much lower sensitivity to diffuse extended emission. We note that at the sensitivity level considered in this work, we expect the number of background sources to be $< 10^{-5}$~arcsec$^{-2}$ \citep{padovani2011}, so there would be no significant contamination in the simulated field of view shown in \cref{fig:ddrg-plot}.

If the source is expanding into a large-scale cluster environment (e.g., n400-Q44), eMERLIN detects three distinct radio components at all three redshifts -- an inner lobe pair or compact structure on $\lesssim 5$~kpc scales, and a more extended lobe/hotspot reaching $\sim \!20$~kpc in each direction. While the emission connecting these components is above the chosen sensitivity limits, surveys with sub-arcsecond resolution but lower surface brightness sensitivity may only detect the brighter hotspot and core components. This resembles a `double-double' radio galaxy, a morphology which is typically attributed to restarted sources \citep[e.g.,][]{schoenmakers2000,nandi2012}. Genuine restarted radio sources are typically identified using a combination of double-double morphology, spectral curvature, and high core prominence \citep[e.g.,][]{jurlin2020}. 

The synchrotron radio spectrum for the n400-Q44 source expanding into a cluster environment is resolved into contributions from the `inner' and `outer' lobe pairs (\cref{fig:ddrg-plot}, upper right). The spectrum of the inner double flattens at low ($\lesssim 500$~MHz) frequencies, typical of a CSS source. The outer lobe pair has a steep spectrum, characteristic of an extended radio source. The absence of high-frequency spectral curvature for the outer lobe pair is consistent with a relatively short-lived remnant phase -- the break frequency is a function of source age and lobe magnetic field \citep[e.g.,][]{brienza2020}, and is expected to lie well above the range of observing frequencies for this source. The curvature of the overall spectrum is due to the relative contributions from each of the lobe pairs, which in turn depends on the surrounding environment. Approximately $20\%$ of the total $1.4$~GHz emission from this source is from the inner double (see Section~\ref{sec:core-lobe}), with candidate restarted sources typically selected if this fraction is above $10\%$ \citep[cf.][]{jurlin2020}.

Alternatively, if the source is expanding into a steeper large-scale environment (n400-Q44-group, shown in the bottom row in \cref{fig:ddrg-plot}), particularly if the source is highly asymmetric due to interactions within the host galaxy (Section~\ref{sec:asymmetry}), a different morphology may be observed. At redshifts of $z = 0.5$ and $z = 3$, eMERLIN detects two main structures -- continuous emission from the host galaxy/inner double and southern lobe, and very faintly connected emission from the northern lobe. Depending on the survey surface brightness sensitivity, or the location of random noise in the image, it may be difficult to conclusively associate the emission from the northern lobe with the rest of the source. If this were the case, this source may be misclassified as a head-tail source with an unconnected northern component, with the morphology caused by lobe asymmetries and the diffuse large-scale environment. The degree of any misclassification will depend on the relative brightness of the connecting bridge, with this difference becoming particularly prominent at a redshift of $z = 3$; this is further quantified in \ref{app:extra-figures2}, \cref{fig:sb-z-ddrg}. The synchrotron radio spectrum of this source (n400-Q44-group) is similar to that expanding into a large-scale cluster environment (n400-Q44), but with the luminosity of the outer lobe pair lower by a factor of two. This results in a greater contribution to the total emission from the inner lobe pair, increasing both the core prominence (see Section~\ref{sec:core-lobe}) and the overall spectral curvature. 

While the simulations performed in this work are idealised, with no cluster-scale environment asymmetries or winds included, we expect the general results to hold on the range of spatial scales considered in this work, potentially even heightened for different combinations of jet and environment properties. For example, a low-power jet (e.g., n400-Q43) will produce lower luminosity extended emission -- if this falls below survey sensitivity limits, the source may be detected as a compact/galaxy-scale structure even though jet material has reached much larger distances. In fact, for jet material extending to a total source size of $25$~kpc, lobe emission is so faint for the low-power source that any emission from outside the host galaxy ($\gtrsim 5$~kpc) is only just detected by eMERLIN, even at a redshift of $z = 0.05$, resulting in heightened core prominence values despite the continuous activity of the source.

\subsection{Environment effects} \label{sec:core-lobe}

\begin{figure}[t]
	\includegraphics[width=1.002\columnwidth,trim={6 7 5 4},clip]{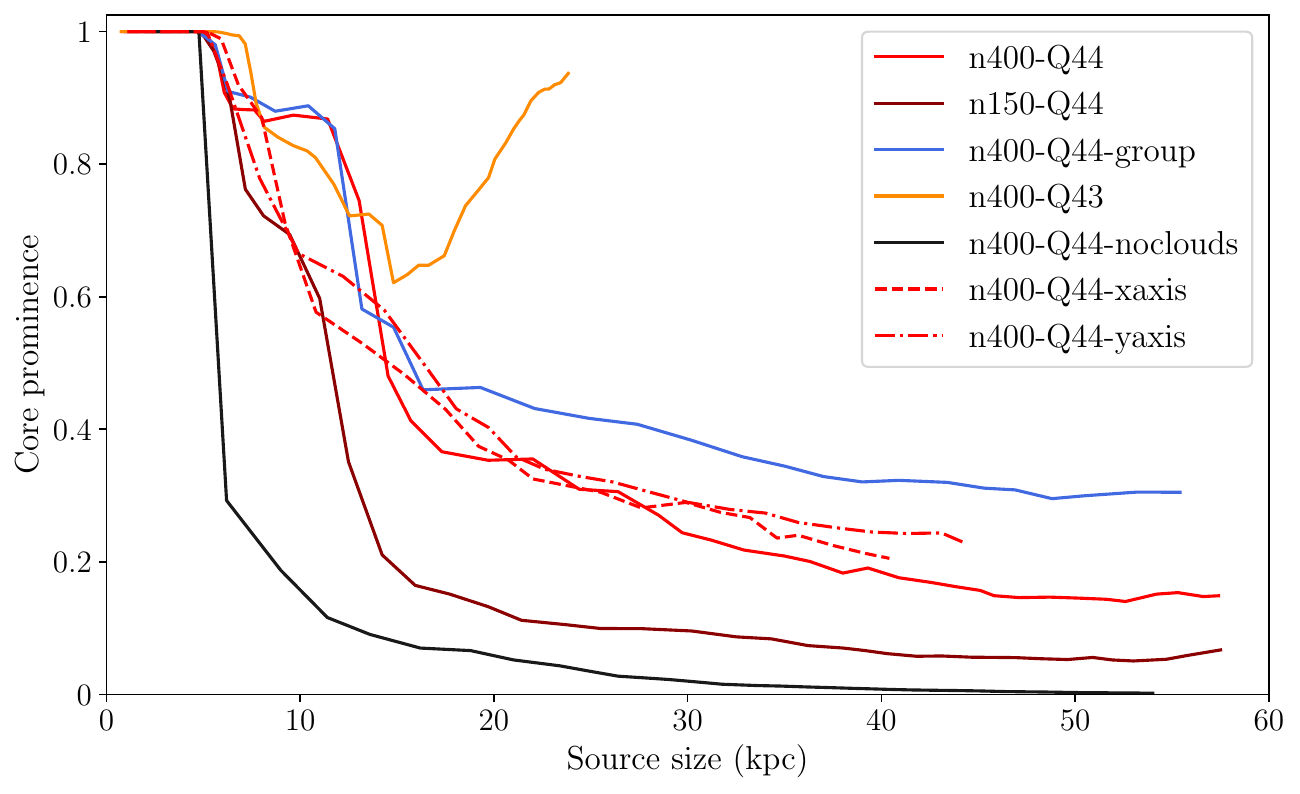}
    \caption{Core prominence, defined as $P_{1.4~\mathrm{GHz}}$($r < 2.5$~kpc) / $P_{1.4~\mathrm{GHz}}$(whole source), for our simulations as a function of their total source size. }
    \label{fig:core-prominence}
\end{figure}

The core prominence (CP), defined here as the fraction of the total $1.4$~GHz radio power from within the host galaxy ($r < 2.5$~kpc), is frequently invoked to associate core emission as part of a more recent outburst than the extended emission, and to classify remnant radio galaxies \citep[e.g.,][]{hardcastle2016,brienza2017,jurlin2020}. We examine the contributions of emission from the host galaxy to the total emission in \cref{fig:core-prominence} for each simulated source. This value is strongly related to the ratio of the host galaxy density to that of the larger scale environment; we note that the host galaxy dominates the total emission for our simulations at sizes below $15$~kpc.

\begin{figure*}
    \includegraphics[width=\textwidth]{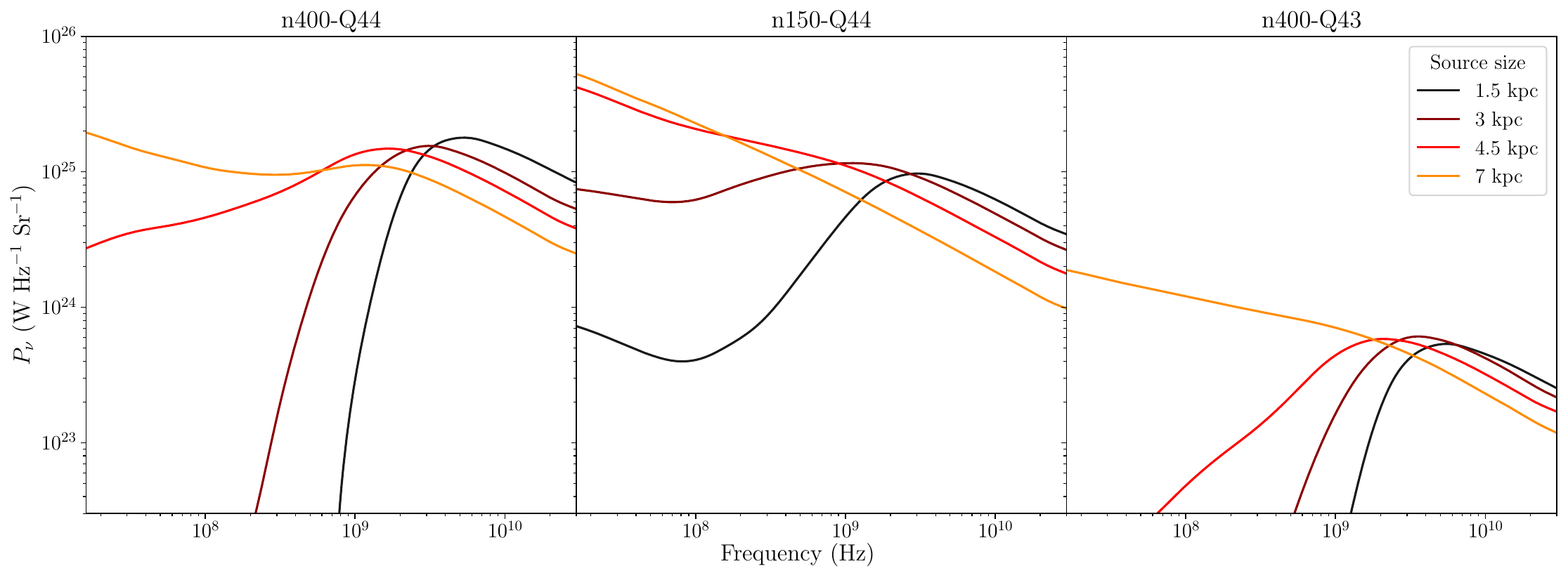}
    \caption{Example synchrotron radio spectra for (L-R) n400-Q44, n150-Q44, and n400-Q43 simulations. Spectra are shown for total source sizes of $1.5$, $3$, $4.5$, and $7$~kpc.}
    \label{fig:SEDs}
\end{figure*}

For the same large-scale cluster environment (i.e., n400-Q44, n150-Q44, and n400-Q44-noclouds), brighter core emission is associated with higher density clouds in the inner regions of the galaxy. This is due to two main contributing factors: (i) individual high density clouds cause splitting or branching of the jet, resulting in multiple sites of shock acceleration within the host galaxy, with emitting particles distributed over a much wider region; and (ii) since the source takes longer to expand through a denser ISM, more energy has been injected by the jet for the same source size. Both of these factors are clearest for simulation n400-Q43 (low jet power), since this source experiences slower source expansion and has a greater density contrast between jet and cloud material than higher-power sources, resulting in the jet being more easily deflected through low-density channels. The observed increase in core prominence for the low-power jet at source sizes above $15$~kpc is due to the combination of decreasing lobe brightness caused by volume expansion and adiabatic losses, and increasing core brightness due to continued lateral expansion of the jet within the galaxy.

As distance from the jet axis increases, we find that there is a decrease in the shock age of particles outside the jet channel. That is to say, the most recently shocked particles are located furthest from this axis where they are interacting with dense clouds. Older particles tend to `bunch up' behind this region, with the oldest particles experiencing little direct interaction with their surroundings. We also find that there is a corresponding increase in luminosity as the jet first impacts a particularly dense cloud (such as those investigated in Section~\ref{sec:asymmetry-genesis}), so we expect a greater contribution to the overall luminosity from within the host galaxy while the jet is still branching out and interacting with new dense clouds.

For the same host galaxy but different large-scale environment (i.e., n400-Q44 and n400-Q44-group), the lower density large-scale group environment results in fainter \emph{lobe} emission, and thus a higher fraction of the total emission will be from within the host galaxy. This also results in greater curvature of the synchrotron radio spectrum, since the free-free absorbed host galaxy emission is more dominant (see \cref{fig:ddrg-plot}).

The range of core prominence values here are significantly higher than those observed for $100$-kpc scale active or remnant sources. Active sources are typically observed with a core prominence of $\sim\!0.001$--$0.1$ \citep{deruiter1990, mullin2008}, while remnant sources have CP $\lesssim 10^{-3}$ \citep[e.g.,][]{hardcastle2016,brienza2017}. Restarted sources, meanwhile, generally have high core prominence values \citep[CP~$> 0.1$;][]{jurlin2020}. The relatively high core prominence values of our simulated radio galaxies are due to the young ages of the simulated sources, which have not yet inflated luminous, large-scale lobes.

\section{Dependence of radio spectra on jet and environment properties} \label{sec:seds}

The impact of free-free absorption (FFA) on the synchrotron radio spectrum depends on the properties of the host galaxy environment. \cref{fig:SEDs} shows the synchrotron radio spectra for each of the n400-Q44, n150-Q44, and n400-Q43 simulations at a range of source sizes. As each source grows, the peak in the spectrum shifts to lower frequencies. This is qualitatively consistent with the observed trend of \cite{odea1997}, who found an inverse correlation between turnover frequency ($\nu_\mathrm{p}$) and linear source size (LS) for GPS and CSS sources, with $\nu_\mathrm{p} \propto \mathrm{LS}^{-0.65}$. The effects of ISM density (Section~\ref{sec:seds-ism}) and jet power (Section~\ref{sec:seds-power}) on the evolution of turnover frequency with source size are discussed in the following subsections.

We also highlight the low-frequency upturn in the spectra of simulation n150-Q44 while the source is still within the host galaxy. These spectra are remarkably similar to those identified by \citet[][see their Figure~10]{callingham2017}, with convex spectra suggested as another signpost of multiple epochs of AGN activity. We propose that this may instead be due to the partial screening of the jets by the absorbing clouds, with our spectra representing a combination of a free-free absorbed component which dominates at higher frequencies and an optically thin component which dominates at frequencies below the turnover.

\begin{figure}[t]
    \includegraphics[width=1.002\columnwidth,trim={6 7 5 4},clip]{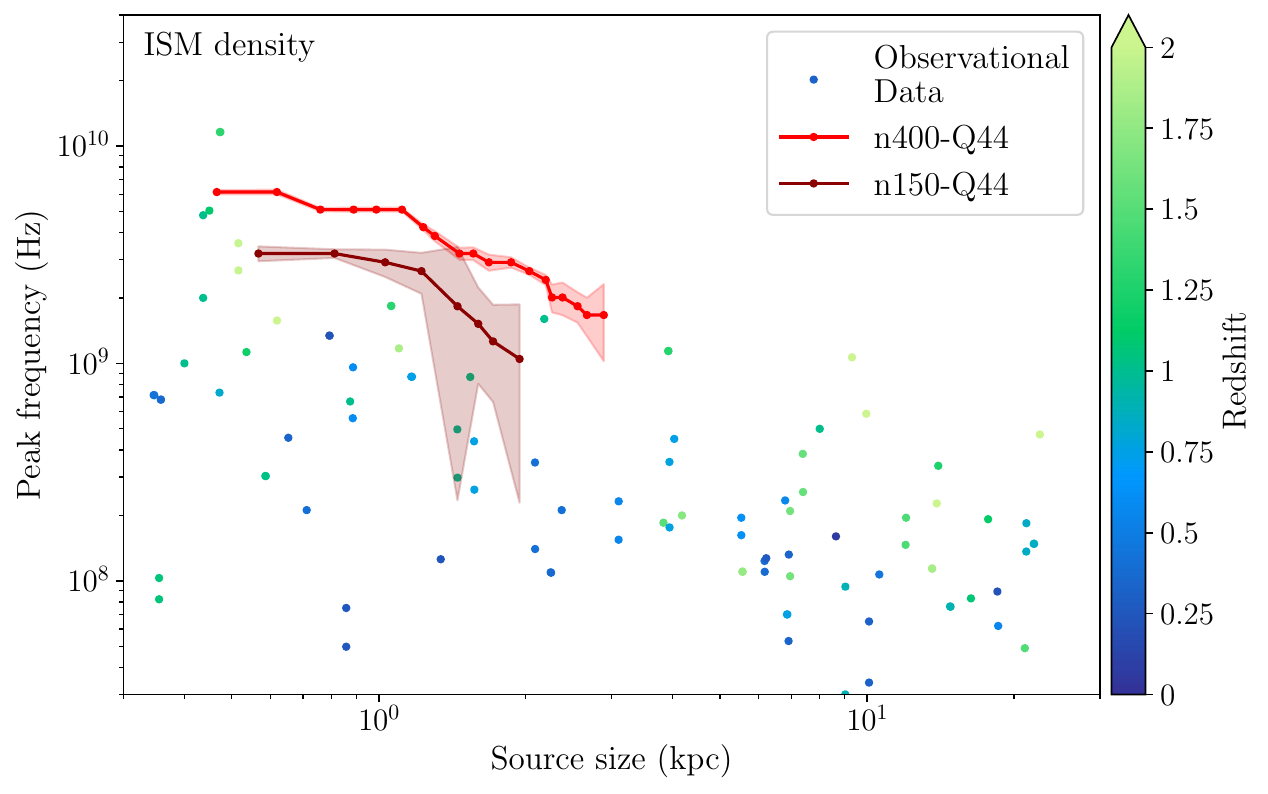}
    \includegraphics[width=1.002\columnwidth,trim={6 7 5 4},clip]{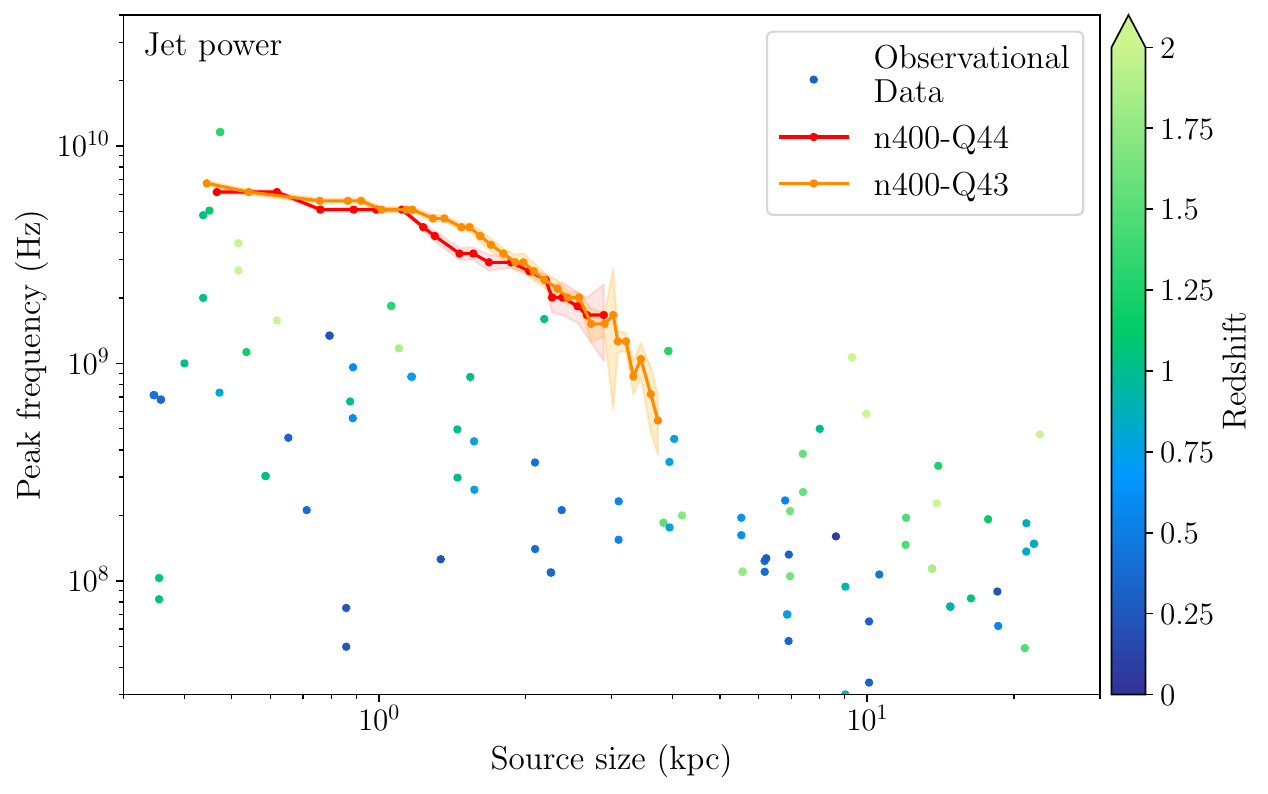}
    \caption{Evolution of (rest-frame) peak frequency with source size for simulations with different ISM densities (top, n400-Q44 and n150-Q44) and jet powers (bottom, n400-Q44 and n400-Q43). Shaded regions represent uncertainty in peak frequency values. Blue points represent the data of \citet{odea1997} and \citet{jeyakumar2016}, coloured by source redshift. Simulation source sizes have been reduced by a factor of $\sqrt{3}$ to account for projection effects for a source at the median viewing angle.}
    \label{fig:nu-LS-150-43}
\end{figure}

\subsection{ISM density} \label{sec:seds-ism}

For the same source size, a lower density host galaxy environment results in the turnover shifting to lower frequencies. The free-free optical depth, $\tau_\nu$, is proportional to the amount of absorbing material along each line of sight to the source (see \cref{eqn:intensity,eqn:abscoeff}). Since there is less absorbing material along a given line of sight to the source for n150-Q44, signatures of FFA have disappeared almost entirely by the time the source has expanded past the edges of the host galaxy (LS $\gtrsim 5$~kpc). 

The trends across the simulated source lifetimes are shown in \cref{fig:nu-LS-150-43}, with the shaded regions representing the uncertainty in modelled peak frequency values, fitted using the model of \citet{callingham2015}. For all source sizes which exhibit a turnover in the synchrotron radio spectrum, the peak frequency is lower for the low ISM density simulation. This is consistent with the results of \cite{bicknell2018}, who investigated this effect for densities ranging from $150$--$2000$~cm$^{-3}$. 

\cref{fig:nu-LS-150-43} shows that our simulated jets, particularly those with a dense ISM (n400-Q44 and n400-Q43), lie towards the higher-frequency end of the observational data from \citet{odea1997} and \citet{jeyakumar2016}. This suggests that an ISM density of $150$~cm$^{-3}$ is perhaps more representative of the environments surrounding observed sources than higher densities, however this relationship is likely to also depend on the scales over which the clouds are distributed. The maximum extent of clouds encountered by each jet in our simulations is $2.5$~kpc; this limits the largest source sizes showing a free-free turnover to a few kpc, well below the largest sizes of observed peaked-spectrum sources. Larger simulated sources with FFA signatures will be possible for more realistic gas--jet geometries, such as jets which are inclined with respect to the galactic disk \citep[e.g.,][]{mukherjee2018}. We defer detailed analysis of alternative cloud distributions to future work. We note that the simulations which have the same \emph{average} ISM density but a different arrangement of dense clouds (i.e., n400-Q44-xaxis and n400-Q44-yaxis) evolve in the same way through $\nu_\mathrm{p}$--LS phase space.

\subsection{Jet power} \label{sec:seds-power}

Despite some differences in how the low-power jet interacts with the clouds in its vicinity (see Section~\ref{sec:asymmetry-genesis}), the behaviour of the peak frequency with source size is similar for the n400-Q44 and n400-Q43 simulations. Due to the lower energy being injected via the jet, we expect this source to consistently have a lower luminosity than its high-power counterpart (as seen in \cref{fig:SEDs}), resulting in a downward shift in normalisation of the synchrotron radio spectrum. The free-free absorption coefficient is dependent on environment properties such as electron density and temperature, and the frequency of emission being absorbed. There is no explicit dependence of the free-free absorption coefficient on the properties of the jet, hence the turnover frequency and general shape of the spectrum do not change with jet power.

The evolution of n400-Q44 and n400-Q43 through $\nu_{\rm p}$--LS phase space are consistent within spectral fitting uncertainties (\cref{fig:nu-LS-150-43}). This result is in broad agreement with the simulations of \cite{bicknell2018}, although these authors did not explicitly mention the lack of dependence on jet power in their work. This provides a potential method of `weighing' the ISM surrounding a young radio source: if the source size and rest frame turnover frequency can be measured for a given source, the jet power does not need to be known to make an estimate of the central gas density.

\section{Conclusions} \label{sec:conclusion}

In this paper, we have investigated the role of galactic and larger-scale environment properties on the evolution of young radio sources. We performed hydrodynamic simulations of AGN jets interacting with a dense multiphase ISM and a larger-scale diffuse environment which, for the first time, include both free-free absorption and a semi-analytic treatment of full radiative and adiabatic losses of synchrotron-emitting electrons. Our main results are as follows:

\begin{enumerate}[label=(\roman*)]
    \item Asymmetries in both lobe length and synchrotron brightness develop due to ISM inhomogeneities. Jet--cloud interactions are influenced both by the mass and density of individual clouds, and the ram pressure of the jet upon impact. The magnitude of the asymmetries depends on how much the jet and counterjet are held up by individual clouds before reaching the edge of the host galaxy.
    \item Asymmetries developed in the galaxy persist to greater radii in rapidly declining large-scale environments such as poor groups, than in rich environments such as clusters.
    \item Morphology of young radio sources may depend on resolution and surface brightness sensitivity of observations. Specifically, apparent `double-double' radio morphologies may be produced without any jet intermittency.
    \item The fraction of total radio emission from the core depends on the relative densities of the host galaxy and circumgalactic gas. For the same large-scale environment and jet power, higher ISM density results in branching of the jet, producing multiple sites of shock acceleration within the host galaxy and a brighter core. Low density larger-scale environments produce lobes of lower luminosity, also increasing core prominence. 
    \item The young active sources (LS~$\lesssim 60$~kpc) examined in our work exhibit core prominence values above the cutoff typically used to classify restarted radio sources (CP~$> 0.1$). We propose that additional remnant/restarted source classification metrics be applied in conjunction with core prominence to robustly classify these objects.
    \item At a fixed source size, the peak frequency is lower for a lower density ISM. Because peak frequency is independent of jet power, radio spectra towards free-free absorbed sources may provide a new method for measuring the ISM.
\end{enumerate}

\paragraph{Acknowledgments}
We thank the anonymous referee for a helpful and constructive report which improved this paper. SY and GS thank the University of Tasmania for an Australian Government Research Training Program (RTP) Scholarship. This research was carried out using the high-performance computing facilities provided by Digital Research Services, IT Services at the University of Tasmania. We acknowledge the work and support of the developers providing the following \textsc{python} packages: \textsc{astropy} \citep{astropy2022}, \textsc{jupyterlab} \citep{jupyterlab2016}, \textsc{matplotlib} \citep{matplotlib2007}, \textsc{numpy} \citep{numpy2020}, and \textsc{scipy} \citep{scipy2020}. 

\paragraph{Funding Statement}
This research was supported by the Australian Research Council via grant DP240102970 (SS and PYJ).

\paragraph{Conflicts of Interest}
None

\paragraph{Data Availability Statement}
The data underlying this article will be shared on reasonable request to the corresponding author.

% \paragraph{Ethical Standards}
% The research meets all ethical guidelines, including adherence to the legal requirements of the study country.

\paragraph{Author Contributions}
Conceptualisation: R.J.T; S.S.S. Formal analysis: S.A.Y. Methodology: S.A.Y; R.J.T; S.S.S.; G.S.C.S. Software: S.A.Y; G.S.C.S; P.M.Y-J. Visualisation: S.A.Y. Writing -- original draft: S.A.Y. Writing -- review \& editing: S.A.Y; R.J.T; S.S.S. All authors approved the final submitted draft.

%\endnote in some journals will behave like \footnote; and \printendnotes will not output anything. 
\printendnotes

\bibliography{example}
\newpage\onecolumn
\appendix

\section{Jet--cloud interactions -- individual clouds} \label{app:extra-figures}

\begin{figure*}
    \includegraphics[width=\textwidth]{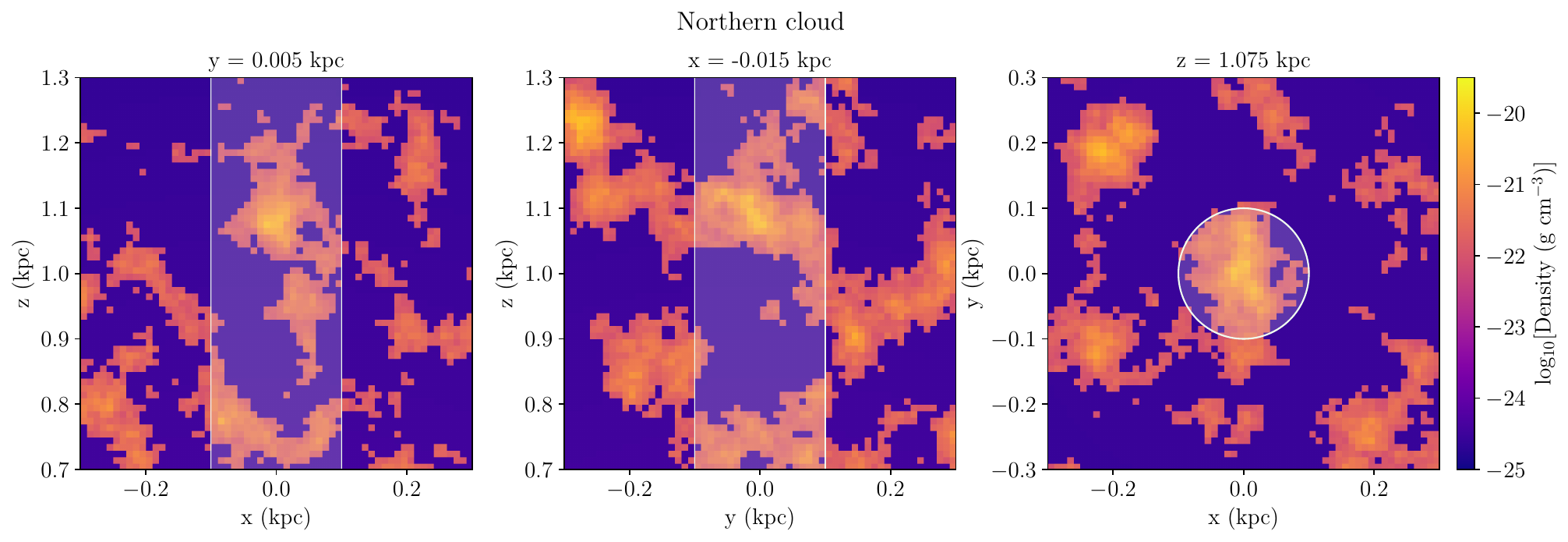}
    \includegraphics[width=\textwidth]{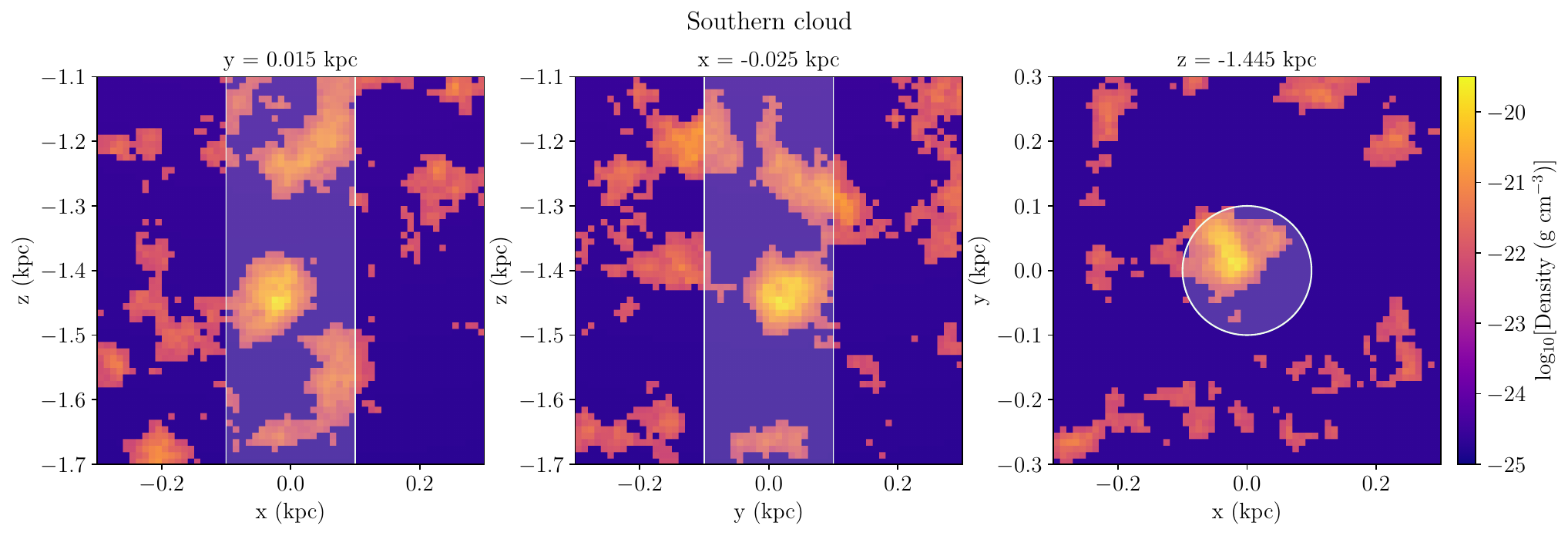}
    \caption{Midplane density slices in the $xz$ (left), $yz$ (middle), and $xy$ (right) planes for the northern (top row) and southern (bottom row) clouds at $t=0$, overlaid with a jet cylinder of radius $0.1$~kpc.}
    \label{fig:cloud-slices-combined}
\end{figure*}

We present midplane density slices for the two dense clouds discussed in Section~\ref{sec:asymmetry-genesis}. The upper row of \cref{fig:cloud-slices-combined} shows the `northern' dense cloud, located at $(x, y, z) = (-0.015, 0.005, 1.075)$~kpc. The jet cylinder is highlighted in white, with the jet travelling from bottom to top in the left and middle panels. The second row of \cref{fig:cloud-slices-combined} shows the `southern' dense cloud, located at $(x, y, z) = (-0.025, 0.015, -1.445)$~kpc. The jet travels from top to bottom in the left and middle panels. The northern cloud occupies a larger fraction of the jet cylinder, but has a lower peak density.

\newpage
\section{Misclassification of active sources -- brightness of connecting features} \label{app:extra-figures2}

\begin{figure}
    \includegraphics[width=0.5\columnwidth]{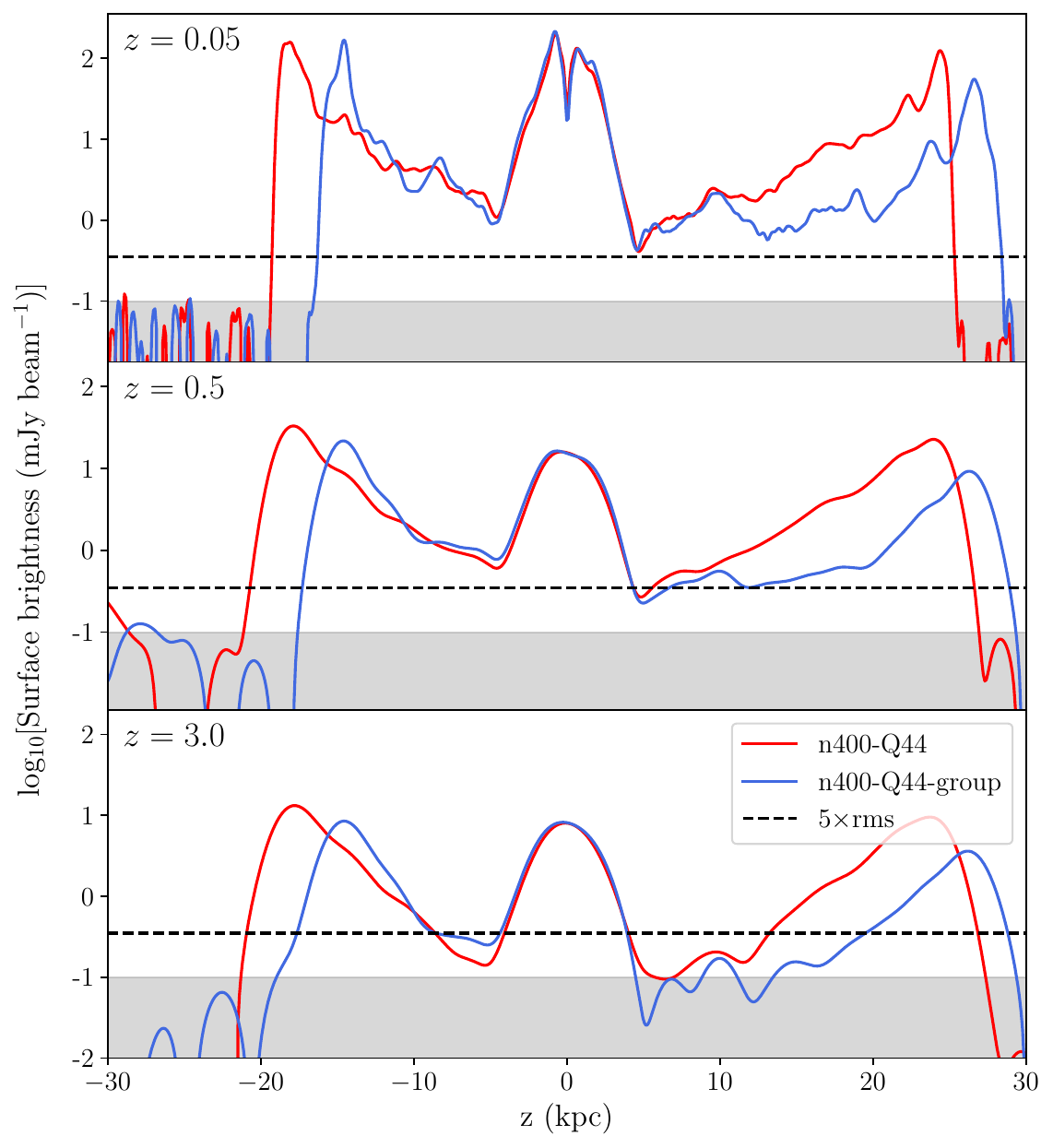}
    \caption{Surface brightness along the jet axis ($x = 0$) for the sources shown in \cref{fig:ddrg-plot} at redshifts of $z = 0.05$ (top), $z = 0.5$ (middle), and $z = 3$ (bottom). The black dashed line is at five times the rms noise placed on the surface brightness images, with rms $= 70 \, \mu$Jy~beam$^{-1}$ \citep[cf. median rms from LeMMINGs;][]{lemmings2018}. The grey shaded region represents features below the chosen sensitivity of $0.1$~mJy~beam$^{-1}$.}
    \label{fig:sb-z-ddrg}
\end{figure}

In Section~\ref{sec:morphologies}, we discussed the potential misclassification of young radio sources depending on the relative surface brightness of the bright components (such as the core and hotspots) compared to that of the faint bridge of emission connecting them. In \cref{fig:sb-z-ddrg}, we quantify this difference for the sources shown in \cref{fig:ddrg-plot} by plotting the surface brightness along the jet axis ($x = 0$). At redshifts of $0.05$ and $0.5$, all emission above the noise level lies above the detection limit of $0.1$~mJy~beam$^{-1}$, representative of the sensitivity of eMERLIN \citep{lemmings2018}. For the source expanding into a large-scale group environment at $z = 0.5$, the bridge connecting the core and the northern (z $> 0$~kpc) hotspot has surface brightness comparable to five times the rms noise level ($\sim 0.35$~mJy~beam$^{-1}$). By contrast, at a redshift of $3$, all of the connecting features drop below this level, with the separation between `bright' features reaching approximately $15$~kpc. Hence at $z = 3$ this source is likely to be identified as a collection of distinct components, rather than as a single structure. 

\end{document}